\documentclass[aps,prd,reprint,superscriptaddress,nofootinbib]{revtex4-2}
\usepackage{amsmath,amsfonts,amssymb,bm}
\usepackage{graphicx}
\usepackage{bbold}
\usepackage[normalem]{ulem}
\usepackage[dvipsnames,usenames]{xcolor}
\usepackage[colorlinks,citecolor=blue]{hyperref}
\usepackage{orcidlink}

\begin{document}

\title{Non-Equilibrium Steady States in Nucleosynthesis}

\author{Zewei Xiong\orcidlink{0000-0002-2385-6771}}
\email[Email: ]{z.xiong@gsi.de}
\affiliation{GSI Helmholtzzentrum {f\"ur} Schwerionenforschung, Planckstra{\ss}e 1, 64291 Darmstadt, Germany}

\author{Gabriel Mart{\'i}nez-Pinedo\orcidlink{0000-0002-3825-0131}}
\affiliation{GSI Helmholtzzentrum {f\"ur} Schwerionenforschung, Planckstra{\ss}e 1, 64291 Darmstadt, Germany} \affiliation{Institut f{\"u}r Kernphysik (Theoriezentrum), Fachbereich Physik, Technische Universit{\"a}t Darmstadt, Schlossgartenstra{\ss}e 2, 64289 Darmstadt, Germany}

\date{\today}

\begin{abstract}
Steady-state behavior underlies many nucleosynthesis processes in which the population number of nuclei can remain nearly constant despite the rapid nuclear reactions.
While existing theoretical descriptions largely focus on established chemical equilibrium, as in nuclear statistical equilibrium (NSE) and quasi equilibrium (QSE), non-equilibrium steady states (NESSs) can form in process-specific approximations such as classical $\beta$-flow $r$-process.
We develop a general framework for describing NESSs in nuclear reaction networks by formulating the steady-state condition as a balance of reaction flows within a reduced linear description of the relevant network subset.
We apply the framework to four representative nucleosynthesis scenarios spanning different steady-state regimes: disequilibrium between nucleons and $\alpha$ particles during $\alpha$-rich freeze-out, the classical $\beta$-flow steady state, the neutrino-assisted $\nu r$-process, and the fission-cycling phase of the $r$-process.
In particular, the fission-cycling scenario is described by a quasi-steady-state (QSS) approximation maintaining the population ratios between two nuclei while not conserving the total number of nuclei.
The framework reproduces the characteristic abundance pattern for each case while providing a unified interpretation of the underlying steady-state behavior.
The eigenvalue spectrum reveals network structure and characterizes the relaxation timescales of abundance distributions.
These results establish a general approach for analyzing nucleosynthesis beyond the condition of chemical equilibrium and provide a foundation for extending graph-based methods to complicated nuclear reaction networks.
\end{abstract}

\maketitle
\graphicspath{{./}{figures/}}

\section{Introduction}
\label{sec:introduction}
Nucleosynthesis is the fundamental processes responsible for the production of chemical elements through a network of nuclear reactions operating in a wide range of astrophysical environments.
These reactions synthesize nuclei from free nucleons and lighter seed nuclei, thereby shaping the evolution of stars and the chemical enrichment of galaxies.
The abundance distribution produced in a specific astrophysical environment is determined by the interplay between nuclear reactions and the thermodynamic evolution of the astrophysical environment.
Depending on the temperature, density, and neutron-to-proton ratio, different reaction mechanisms dominate, giving rise to distinct nucleosynthesis processes such as stellar burning, slow ($s$-) and rapid neutron-capture ($r$-) processes \cite{burbidge1957synthesis,Cameron:1957,*Cameron:2013,meyer1994pprocesses,cowan2021origin,thielemann2026rprocess}.

In many nucleosynthesis environments, nuclear reactions proceed much faster than the hydrodynamic evolution of the expanding material while the populations of individual nuclear species change only slowly.
This behavior arises when the production and destruction of each species approximately balance, even though individual reaction channels remain active and carry non-negligible fluxes between two nuclear species.
The resulting steady state reflects a dynamical equilibrium among competing nuclear reactions and persists over timescales much longer than the underlying reaction scales.

A typical example is nuclear statistical equilibrium (NSE), in which each forward reaction is balanced by its inverse reaction under sufficiently high temperatures and densities.
As a consequence, the nuclear abundances are uniquely determined by thermodynamic equilibrium and can be calculated from the temperature, density, electron fraction, and nuclear properties without explicit knowledge of the individual reaction rates~\cite{hoyle1960nucleosynthesis}.
NSE provides an accurate description of matter in environments such as the cores of core-collapse supernovae and merging neutron stars, where the reaction timescales are predominantly shorter than the dynamical timescale.

As the temperature and density decrease, the conditions required for global equilibrium gradually break down because some reactions become too slow.
Nevertheless, groups of nuclei connected by sufficiently rapid reactions may continue to satisfy local equilibrium conditions, leading to quasi-statistical equilibrium (QSE).
Instead of a single equilibrium involving all nuclei, the reaction network separates into one or more QSE clusters whose internal abundance ratios remain close to their NSE values while the exchange of material between different clusters is controlled by comparatively slow bottleneck reactions \cite{bodansky1968nuclear,meyer1998theory,hix1999silicon,hix2007qsereduced}.
The QSE approximation therefore retains the dominant equilibrium while substantially reducing the complexity of the full reaction network.

Despite their success, NSE and QSE describe only a limited range of conditions relevant to nucleosynthesis.
As the thermodynamic conditions evolve, the reaction network may no longer satisfy the equilibrium relations required for either global or local equilibrium.
Nevertheless, the abundance distribution does not necessarily become completely arbitrary once chemical equilibrium is no longer maintained.
In many astrophysical environments, nuclear populations can still reach a dynamical steady state even when equilibrium between forward and backward reactions breaks down.
Such a state is referred to as a non-equilibrium steady state (NESS), in which the abundances remain stationary under the collective action of the reaction network despite the absence of equilibrium constraints.

The evolution of a nuclear reaction network is fundamentally governed by its connectivity.
Each isotope may be regarded as a node in a weighted directed graph, while each nuclear reaction corresponds to a directed edge whose weight is determined by the corresponding reaction rate.
Network theory provides a general mathematical framework for describing such interconnected systems and has found broad applications in the natural sciences as well as in fields such as computer science to finance (see, e.g., \cite{schnakenberg1976network,boccaletti2006complex,seifert2012stochastic,newman2018networks,bardoscia2021physics}).
Many general properties of these systems depend primarily on the structure of the underlying network rather than on the detailed properties of its individual components.
Representing a nuclear reaction network as a weighted directed graph provides a natural framework for formulating its evolution and analyzing the resulting steady-state behavior.
Concepts such as connectivity, conservation laws, steady states, and relaxation modes can then be understood from the network structure, regardless of whether the system satisfies equilibrium conditions.

In nucleosynthesis, NESSs arise when the nuclear abundance does not evolve even though the conditions for chemical equilibrium are not met.
This behavior can occur in both complete and partial out-of-equilibrium evolution, with different groups of reactions satisfying different steady-state conditions.
Examples include $\beta$-flow steady state of the $r$-process, disequilibrium between nucleons and $\alpha$ particles during $\alpha$-rich freeze-out, sufficiently neutron-rich fission cycling, the neutrino-assisted $\nu r$-process, the CNO cycle~\cite{bethe1939energy}, steady neutron-capture flow in $s$-process~\cite{clayton1961neutron}, as well as the transient radioactive equilibrium in decay chains involving fissions and $\beta$ decay, as encountered in nuclear experiments.
We focus on the first four scenarios in this paper.

The most widely studied example is the $\beta$-flow steady state in the classical $r$-process.
The material flow between neighboring isotopic chains becomes nearly constant, regulated by $\beta$ decays of the waiting-point nuclei.
The resulting abundance pattern is therefore determined by the balance between the reaction fluxes into and out of each isotopic chain \cite{cameron1983steady,cameron1983waiting,cowan1991r,kratz1993isotopic,arcones2011dynamical}.

Another example occurs in rapidly expanding material, where the equilibrium between nucleons and $\alpha$ particles can break down, leading to a different abundance pattern and allowing an $r$-process to develop even in the absence of excess free neutrons \cite{meyer2002rprocess}.
In this regime, the reaction network remains far from equilibrium, yet the competing reaction channels establish a steady abundance flow among the synthesized heavy nuclei.

A similar phenomenon has recently been identified in the $\nu r$-process, where neutrino absorption plays the role that photodisintegration or $\beta$ decay plays in counteracting neutron captures in the classical $r$-process.
Although a different combination of reactions drives the abundance evolution, steady-state reaction flows again
determine the abundances in the absence of chemical equilibrium~\cite{xiong2024production}.

The steady state defined above requires the populations of all individual nuclear species to remain constant.
However, the total number of heavy nuclei need not be conserved, for example, in the sufficiently neutron-rich $r$-process where fission cycling becomes important.
A single fission event produces multiple daughter nuclei, increasing the number of heavy nuclei and redistributing material throughout the network.
Consequently, a conventional steady state with constant populations of all nuclear species may not be attainable.
Nevertheless, the system may approach a state in which the ratios between the populations of any two nuclear species remain constant, even as their absolute populations evolve.
We refer to such a state as a quasi-steady state (QSS).
Although the total number of heavy nuclei continues to grow as neutrons are consumed, the relative abundance approaches a QSS after several fission cycles \cite{seeger1965nucleosynthesis,beun2008fission,goriely2015fundamental,mendoza2015nuclear}.
The resulting distribution is maintained by the interplay of neutron captures, $\beta$ decays, and the redistribution of material by fission.

These examples suggest that NESSs and QSSs are general features of nuclear reaction networks rather than isolated phenomena associated with a particular nucleosynthesis process.

In this work, we establish a unified theoretical framework for describing NESSs in nuclear reaction networks.
We show that a wide variety of steady-state phenomena with different manifestations can be understood within the same mathematical formulation, independent of the specific reactions that establish the steady state.
We further develop practical numerical methods for computing steady-state abundance distributions, including spectral decomposition and perturbation theory, and extend the formalism to QSSs in systems that do not conserve total abundance.

This paper is organized as follows.
Section~\ref{sec:NSE} and Section~\ref{sec:QSE} review the formalism of NSE and QSE, respectively.
Section~\ref{sec:NESS} presents the general theory of NESSs and numerical methods for computing the steady-state abundance distribution.
Section~\ref{sec:QSS} extends the formalism to systems without abundance conservation as QSSs.
Section~\ref{sec:results} applies the theory to the aforementioned nucleosynthesis scenarios under fixed thermodynamic conditions, comparing the results with those obtained from full nuclear reaction network calculations.
Finally, Section~\ref{sec:conclusion} summarizes the main results and discusses their implications for nucleosynthesis and future applications.

\section{Nuclear Statistical Equilibrium}
\label{sec:NSE}
For a thermodynamic system undergoing nucleosynthesis and characterized by temperature $T$ and pressure $P$, the infinitesimal change in the internal energy per nucleon $U$ including the contribution of masses is
\begin{equation}
    dU = T dS - P dV + \mu_e dY_e + \sum_i^{\rm nuclear} \mu_i dY_i,
    \label{eq:dU}
\end{equation}
where $S$ is the entropy per nucleon, $V$ is the volume per nucleon, $Y_e$ is the electron fraction, $\mu_e$ is the electron chemical potential, $\mu_i$ is the chemical potential of species $i$, and $Y_i$ is the abundance of species $i$ per nucleon.
For the conditions considered here, the nuclei are non-relativistic and non-degenerate, with chemical potentials given by
\begin{equation}
    \mu_i = m_i c^2 + k_BT \ln \left[ \frac{\rho Y_i}{G_i m_u} \left( \frac{2\pi \hbar^2}{m_i k_B T} \right)^{3/2} \right],
    \label{eq:mu_i}
\end{equation}
where $m_i$ is the nuclear mass, $c$ is the speed of light, $k_B$ is Boltzmann's constant, $\rho$ is the mass density, $G_i$ is the partition function, $m_u$ is the atomic mass unit, and $\hbar$ is the reduced Planck's constant.
Equation~\eqref{eq:mu_i} shows that the abundance $Y_i$ of species $i$ is proportional to $\exp(\mu_i/k_BT)$.

Chemical equilibrium among the participating nuclear species characterizes NSE.
The equilibrium abundances can be obtained by minimizing the internal energy.
Equivalently, for variations that preserve the conserved quantities,
\begin{equation}
    \sum_i \mu^{\rm NSE}_i dY_i = 0
\end{equation}
where $\mu_i^{\rm NSE}$ is the chemical potential of species $i$ in NSE.
A direct implication of this relation is that, for any isotope $(A_i,Z_i) \rightleftarrows N_i n + Z_i p$ with mass number $A_i$, neutron number $N_i$, and proton number $Z_i$ is
\begin{equation}
    \mu^{\rm NSE}_i = N_i \mu_n + Z_i \mu_p,
\end{equation}
with $\mu_n$ and $\mu_p$ free parameters chosen to satisfy the conservation of total mass fraction
\begin{equation}
    \sum_i A_i Y_i = 1,
\end{equation}
and the charge neutrality
\begin{equation}
    \sum_i Z_i Y_i = Y_e.
\end{equation}

\section{Quasi-Equilibrium}
\label{sec:QSE}
One prerequisite of NSE is that every nuclear species simultaneously satisfies the chemical equilibrium condition, so that a single equilibrium is established throughout the entire reaction network.
In practice, however, equilibrium may not be maintained across the entire nuclear chart because some reactions are insufficiently rapid to establish equilibrium between different clusters of the network.
As a result, the isotopes separate into groups that remain internally equilibrated while being out of equilibrium with one another, forming QSE clusters.
For each cluster, denoted by $I$, an additional abundance constraint is introduced,
\begin{equation}
    Y_I - \sum_{i\in I} Y_i = 0.
\end{equation}

For example, when the triple-$\alpha$ reaction or the ($\alpha$,$\alpha$,$n$) reaction is too slow to maintain equilibrium between light particles and heavier nuclei, the heavy nuclei form a separate QSE cluster with total heavy abundance $Y_h$,
\begin{equation}
    Y_h - \sum_{i\neq n,p,\alpha} Y_i = 0.
\end{equation}
In this construction, the light particles $n$, $p$, and $\alpha$ are treated as an external reservoir that sets the environment for the heavy-nuclei cluster.
Within a QSE cluster, isotopes rapidly transform into one another through the exchange or capture of neutrons, protons, and $\alpha$ particles.
If the reactions linking nuclei within the cluster remain in chemical equilibrium, and the light particles mediating these reactions also satisfy the corresponding equilibrium relations, the differences in chemical potentials are given by
\begin{equation}
    \mu^{\rm QSE}_i - \mu^{\rm QSE}_j = (N_i-N_j)\mu_n + (Z_i-Z_j)\mu_p.
\end{equation}
Therefore, the equilibrium relations determine the internal abundance ratios within the cluster, while the absolute abundances still depend on the total cluster abundance.
In this sense, QSE is a local equilibrium approximation within a subset of the full network.

The validity of the QSE approximation depends on a separation of timescales.
Reactions within a cluster must proceed sufficiently rapidly to maintain local equilibrium, whereas the reactions linking different clusters must be comparatively slow.
Under many astrophysical conditions, this separation arises naturally because the reaction network contains relatively slow reactions connecting otherwise strongly connected groups of nuclei, limiting the flow of material between clusters.

\section{Non-Equilibrium Steady State}
\label{sec:NESS}
The preceding discussion demonstrates that QSE provides a powerful simplification by reducing the determination of the abundance distribution to a small number of independent variables while maintaining local chemical equilibrium within each cluster.
Under equilibrium conditions, the abundance distribution is therefore specified by only a few thermodynamic variables such as the nucleonic chemical potentials.
Furthermore, detailed balance requires that the forward and reverse reactions exactly compensate one another, so that the net reaction flux between every pair of isotopes vanishes.

However, when the conditions deviate from equilibrium, reaction flows remain active while the equilibrium relations are no longer satisfied.
In this regime, the abundance distribution can no longer be determined merely from the equilibrium constraints between isotope pairs and must instead be obtained from the full reaction network.
This motivates the formulation of a general theory of NESSs, in which the abundances remain stationary with a vanishing sum of incoming and outgoing reaction fluxes despite the absence of chemical equilibrium.
As discussed in the Introduction, this framework applies to a broad range of nucleosynthesis scenarios, including the breakdown of light-particle equilibrium, the classical $\beta$-flow steady state in the $r$-process, and the interplay between neutron capture and neutrino absorption in the $\nu r$-process.

Consider a weighted directed graph $G=(V,E)$ in which each vertex $i\in V$ corresponds to an isotope species with abundance $Y_i$, and each directed edge $(i,j)\in E$ represents a nuclear reaction from species $i$ to species $j$.
The edge weights are given by the corresponding reaction rates.
The abundance vector $\mathbf y$ is indexed by the vertices of $G$, with the $i$th component equal to the abundance $Y_i$ of the isotope species $i$.
The network flow along each directed edge, referred to as the reaction flux, is given by the product of its weight and the abundance of the source species.

Isotopes in the network can exchange particles with, or react with particles from, an external reservoir containing nucleons, $\alpha$ particles, or neutrinos.
In the present framework, we treat the abundances of the external-reservoir species as fixed and focus on a reduced subset for which the reaction rates are time-independent.
Under these conditions, if each reaction involves only one type of reactants from the subset, the corresponding Laplacian matrix $\mathbf X$\footnote{Here we adopt a different definition with an opposite sign from the conventional graph Laplacian.} is independent of both $\mathbf y$ and time $t$, and the system becomes linear with the master equation
\begin{equation}
    \dot{\mathbf y} = \mathbf{X} \mathbf y.
    \label{eq:ydot=Xy}
\end{equation}
For general nuclear reaction networks, $\mathbf X$ is a Metzler matrix with non-negative off-diagonal elements.
The diagonal elements are non-positive.

If the reaction network satisfies the following constraint that the sum of the elements in the each column of $\mathbf X$ vanishes,
\begin{equation}
    \sum_i X_{ij} = 0,
    \label{eq:sum_X_ij=0}
\end{equation}
the total abundance $\sum_i y_i$ is conserved within the chosen subset, and $\mathbf X$ is the generator of a continuous-time Markov process \cite{norris1997continuous}.
Under this constraint, the determinant of $\mathbf X$ vanishes, so zero is an eigenvalue, and the corresponding steady-state solutions (the fixed point of the network) lie in the null space defined by
\begin{equation}
    \mathbf{X} \mathbf{y} = 0.
    \label{eq:Xy=0}
\end{equation}

If the network includes reactions such as fission that breaks the above constraint, there is no stationary steady-state solutions.
We will discuss this scenario in Sec.~\ref{sec:QSS}.

\subsection{Diagonalization and Spectral Decomposition}
Many approaches such as weighted Jacobi method \cite{young2014iterative} and spectral decomposition can be used to study the network constrained by Eq.~\eqref{eq:sum_X_ij=0} and to determine their steady-state solution.
In this paper, we will focus on the second approach.

For general networks in Eq.~\eqref{eq:ydot=Xy}, the matrix $\mathbf X$ can be diagonalized, and we have
\begin{equation}
    \mathbf X \mathbf v_k = \lambda_k \mathbf v_k,
    \label{eq:Xv=lamv}
\end{equation}
with eigenvalues $\lambda_k$ and the associated eigenvectors $\mathbf v_k$.
The Perron–Frobenius theorem for Metzler matrices guarantees that the eigenvalue with the largest real part is real,
\begin{equation}
    \mathrm{Im}(\lambda_0) = 0 \text{ if }\mathrm{Re}(\lambda_0)\geq \mathrm{Re}(\lambda_1) \geq \cdots \geq \mathrm{Re}(\lambda_k).
    \label{eq:Perron_Frobenius}
\end{equation}

The uniqueness of the steady-state solution depends on the connectivity of the underlying directed graph.
For a strongly connected network\footnote{Such a network is also referred to as irreducible in the literature.}, in which every node is reachable from every other node along a directed path, the Perron–Frobenius theorem implies that the associated Laplacian matrix $\mathbf X$ with the imposed abundance conservation of Eq.~\eqref{eq:sum_X_ij=0} has a one-dimensional null space.
This means that zero is a simple eigenvalue, $\lambda_0=0$, so the the steady-state abundance distribution is unique once normalized. 

All remaining eigenvalues have strictly negative real parts and correspond to transient modes that decay exponentially with time.
Some eigenvalues may have nonzero imaginary parts, corresponding to damped oscillatory modes.
The general time-dependent solution is
\begin{equation}
    \mathbf y(t) = c_0 \mathbf v_0 + \sum_{k=1} c_k \mathbf v_k e^{\lambda_k t},
\end{equation}
where $\mathbf v_0$ is the steady-state eigenvector, and $v_k$ with $k\ge 1$ are the transient modes.
As $t\rightarrow\infty$, all transient modes vanish as $e^{\lambda_k t}\rightarrow 0$, leaving only the steady-state component.
The steady-state solution can therefore be obtained directly from the eigenvector associated with the zero eigenvalue, without explicitly integrating the time-dependent system.

Abundance conservation further implies that
\begin{equation}
    c_0 \sum_i v_{0,i} = \sum_i y_i \text{,~and~}
    \sum_i v_{k,i} = 0
    \label{eq:sum_v}
\end{equation}
for all $k\geq 1$.
Therefore, each decaying eigenmode represents a redistribution of abundance within the network rather than a net creation or destruction of material.
These transient modes transfer abundance among isotopes while leaving the total abundance unchanged, whereas the steady-state distribution alone carries the conserved abundance that can be normalized.
The amplitudes of these modes can be used to assess the validity of the steady-state approximation and estimate the time required to reach steady state, as shown in later sections.

If the network is not strongly connected, there can be more than one eigenvalue of zero.
A simple example is a network that consists of two disconnected strongly connected subnetworks.
In this case, the null space of $\mathbf X$ is two-dimensional, with each subnetwork possessing its own steady-state solution.
The steady-state abundance distribution of the entire network is therefore determined by the initial distribution of abundance between the two subnetworks.

\subsection{Perturbation Theory}
After introducing a numerical method for computing the steady-state solution, we now examine its dependence on the underlying reaction rates.
This dependence is of particular interest when the steady state cannot be described by chemical equilibrium.
When the conditions supporting QSE are no longer met, changes in the reaction rates perturb the steady-state abundance distribution.
To quantify this response, we employ first-order perturbation theory, yielding a linear sensitivity relation for the abundance response.

Consider a perturbation of the transition matrix, $\mathbf X \rightarrow \mathbf X + \Delta\alpha \mathbf C$, where $\Delta\alpha$ is a small parameter and $\mathbf C$ represents the corresponding perturbative change of the reaction network.
In the following discussion, the external boundary conditions and the normalization constraint are assumed to remain fixed during the perturbation.
The steady-state solution is perturbed accordingly, $\mathbf y \rightarrow \mathbf y + \Delta \mathbf y$.
Substituting these expressions into Eq.~\eqref{eq:Xy=0} and retaining only first-order terms gives
\begin{equation}
    \mathbf X \mathbf y' = \mathbf X \frac{\Delta \mathbf y}{\Delta\alpha} = -\mathbf C \mathbf y,
    \label{eq:Xdyda=-Cy}
\end{equation}
where $\mathbf C \mathbf y$ is the forcing term, and $\mathbf y' \equiv \Delta \mathbf y/\Delta\alpha$ is the abundance response.
This equation determines the sensitivity of the steady-state abundances to changes in the reaction rates.

Formally, if $\mathbf X$ were invertible, the perturbation could be obtained through $\mathbf X^{-1}$, which acts as the Green's function of the network and describes how a local perturbation propagates through the entire reaction network.
However, due to abundance conservation, $\mathbf X$ is singular, and we have
\begin{equation}
    \sum_{i,j} C_{i,j} y_j = 0.
\end{equation}
Because $\mathbf X$ is singular, Eq.~\eqref{eq:Xdyda=-Cy} does not determine a unique $\mathbf y'$.
A unique solution can be obtained by imposing an additional normalization constraint and increasing the matrix dimension.
In practice, the system can also be solve in an alternative way without increasing matrix dimension.
This alternative replaces an arbitrary row of $\mathbf X$ with a row of ones and setting the corresponding entry of $-\mathbf C \mathbf y$ to zero, thereby enforcing the normalization constraint and removing the singularity of the linear system.

Perturbation theory provides a systematic tool for understanding how the reaction network responds to changes in individual reaction rates.
Potentially, this can be used for sensitivity studies as it constitutes a tool to identify the reaction channels that contribute most significantly to the network response.

\subsection{Abundance Conservation}
Both QSE and NESS partition the full reaction network into a subset of interest and an external environment.
Once the abundances outside the subset are specified, the abundance distribution within the subset is determined, and the absolute abundances follow from the conservation of the total abundance.
In both cases, the external network provides the boundary conditions for determining the internal abundance distribution.

QSE can be regarded as a special class of steady state in which every reaction satisfies chemical equilibrium and the net reaction flux between every pair of nuclei vanishes.
In contrast, a NESS allows persistent reaction fluxes while maintaining stationary abundances.
The subset may therefore continuously exchange light particles or other species with the external environment while preserving its internal abundance distribution.
Moreover, weak interactions such as $\beta$ decay can modify the electron fraction $Y_e$ by exchanging the lepton number with the surrounding matter.
Consequently, although the subset itself remains stationary, the external environment may continue to evolve.
If this evolution is too rapid, the conditions required to sustain the steady state are no longer maintained.
Conversely, when the external reservoir is sufficiently large, it provides nearly constant boundary conditions to persist the steady state.

The two descriptions also differ in how the abundance distribution is specified.
In QSE, the distribution in each cluster is fully determined once the density $\rho$, temperature $T$, electron fraction $Y_e$, and the total abundance of that cluster are given, since these quantities determine the neutron and proton abundances entering the equilibrium relations.
In a NESS, however, the abundance distribution is no longer prescribed directly by thermodynamic variables.
Instead, it is determined by the abundances in the external environment together with the reaction fluxes that couple the subset to the rest of the network.
During the evolution toward a steady state, the external abundances and quantities such as $Y_e$ generally coevolve with the subset until a self-consistent stationary abundance distribution is established.

However, the existence of non-zero reaction fluxes does not guarantee that a steady state will be reached.
If the external conditions continue to evolve, the system may remain intrinsically time dependent.
In particular, fluctuations within certain regions of the network can produce wave-like propagation of fluxes when the dominant cycling pathways follow a simple network topology.

\section{Quasi-Steady State}
\label{sec:QSS}
The abundance conservation discussed above depends on the reactions considered in the network.
In certain situations, such as a sufficiently neutron-rich $r$-process that has proceeded through several fission cycle, the total abundance is no longer conserved because fission fragments redistribute the heavy-element population.
The conservation constraint must be modified accordingly.
The diagonalization in Eq.~\eqref{eq:Xv=lamv} still applies, but the dominant eigenvalue can now be nonzero, while the condition in Eq.~\eqref{eq:Perron_Frobenius} still holds.
If fission is the only reaction breaking the abundance conservation, $\lambda_0$ is a positive constant that characterizes the dominant rate of fission cycling.
In this case, the system approaches a QSS, with a fixed relative abundance distribution and a time-dependent total abundance.
The relative abundance distribution can be obtained by
\begin{equation}
    \mathbf y(t) =  e^{\lambda_0 t} \left[ c_0 \mathbf v_0 + \sum_{k=1} c_k \mathbf v_k e^{(\lambda_k-\lambda_0) t}\right],
\end{equation}
where $\lambda_0$ is the dominant eigenvalue and $\mathrm{Re}(\lambda_k-\lambda_0)<0$ for the remaining eigenvalues with $k\geq 1$.

For fission cycling to establish a QSS, the $r$-process must proceed under sufficiently neutron-rich conditions.
A large neutron supply is required to smooth out the transient traffic-wave structure that arises as nuclei are transported from the $\beta$-flow steady state toward the fission region and subsequently redistributed by fission.
If the neutron supply is insufficient, most of the abundance may reach the fission region only once and undergo little subsequent fission cycling, or the nuclei produced during the second cycle may fail to reach the fission region before the neutrons are exhausted.
Consequently, the abundances of individual nuclei continue to evolve from one cycle to the next rather than converging to a fixed distribution.
This behavior manifests itself as oscillatory variations in the abundances of individual nuclei.
As the neutron richness increases, the oscillation amplitudes are damped until the abundance distribution converges to a QSS~\cite{beun2008fission,goriely2015fundamental}.

In the QSS regime, the relative abundance distribution remains unchanged while the total heavy-element abundance grows exponentially.
Consequently, the total heavy mass fraction
continues to increase as heavy nuclei repeatedly consume neutrons supplied by the surrounding environment.

\section{Results in Fixed Thermodynamic Conditions}
\label{sec:results}
The introduction presented four representative nucleosynthesis scenarios that motivate the need for a general theory of NESS.
In this section, we revisit each scenario by computing the steady-state solution from Eq.~\eqref{eq:Xv=lamv} and showing how the formalism provides a unified framework for analyzing the results.

\subsection{Disequilibrium among Nucleons and $\alpha$ Particles}
In a rapidly expanding environment, substantial amounts of free neutrons and $\alpha$ particles may remain after charged-particle freeze-out.
Particularly when the electron fraction is close to $Y_e \approx 0.5$, the conditions favor an $\alpha$-rich freeze-out, with large abundances of $\alpha$ particles and either neutrons or protons.
At sufficiently high temperatures, rapid charged-particle reactions maintain chemical equilibrium among free neutrons, protons, and $\alpha$ particles, allowing the abundances of heavy nuclei to be described by statistical equilibrium.
However, if the expansion is extremely rapid, the expansion eventually outpaces the reactions maintaining equilibrium between $\alpha$ particles and free nucleons, allowing substantial amounts of free neutrons, protons, and $\alpha$ particles to coexist \cite{meyer2002rprocess}.
This causes the heavy-nucleus distribution to depart from QSE.
The resulting out-of-QSE disequilibrium evolution strongly influences the synthesis of heavy nuclei and provides a scenario for applying the NESS formalism.

The NESS calculations consider a heavy-nuclei cluster including all isotopes with $10<N<100$ and $10<Z<70$.
We include strong and electromagnetic reactions induced by photons or light particles, including the $(i,j)$ channels with $i,j\in \{\gamma,n,p,\alpha \}$ and $i\neq j$.
No weak interactions are included.
The parameters that we adopt are $\rho=10^6\,\mathrm{g/cm}^3$, $T=5.72$~GK, $Y_n=0.1$, and $Y_\alpha\approx 0.173$.
The value of $Y_\alpha$ is obtained as the equilibrium abundance value corresponding to $Y_n=0.1$ and $Y_p=10^{-4}$.
We then increase $Y_p$ from $10^{-4}$ to $10^{-3}$, $10^{-2}$, and $0.03$, while keeping $Y_\alpha$ and $Y_n$ fixed, to drive the system progressively away from the QSE distribution.

The network can be constructed with reaction rates determined by the condition given above.
We solve the zero eigenvalue for the Laplacian matrix $\mathbf X$ in Eq.~\eqref{eq:Xv=lamv} and obtain the corresponding eigenvector as the steady-state distribution of heavy nuclei.
We normalize the abundance of heavy nuclei to ensure the total mass fraction to be unity when summed together with abundances of light particles.
This procedure is also applied for all other three scenarios discussed in the following subsections.

For the four cases with $Y_p$ being from $10^{-4}$ to $0.03$, the electron fractions $Y_e$ including both heavy nuclei and light particles are 0.434, 0.434, 0.437, and 0.448, respectively.
The entropy in all cases is in general about $63~k_B/$nucleon.
Compared to the examples in Ref.~\cite{meyer2002rprocess}, the conditions considered here are moderately more neutron-rich and have lower entropy.

\begin{figure}[htb]
    \centering
    \includegraphics[width=\columnwidth]{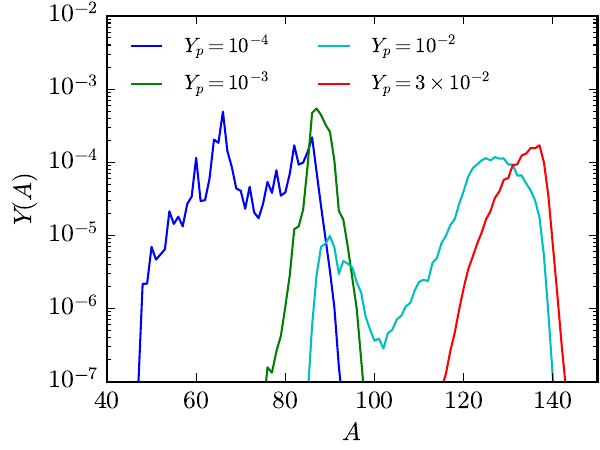}
    \caption{NESS solutions of abundance distributions as a function of mass number $A$ under the nucleon-alpha disequilibrium for four values of $Y_p$. The remaining parameters are given by $\rho=10^6\,\mathrm{g/cm}^3$, $T=5.72$~GK, $Y_n=0.1$, and $Y_\alpha\approx 0.173$.}
    \label{fig:sec1_YA}
\end{figure}

Figure~\ref{fig:sec1_YA} shows the determined NESS solutions of abundance distributions as a function of mass number for the four cases.
When $Y_p=10^{-4}$, the majority of nuclei has mass number $A$ spanning from $\sim 50$ to $\sim 90$.
As we increase the proton abundance, the distribution is shifted drastically from the one peaking at $A\sim 85$ at $Y_p=10^{-3}$ to the one peaking at $A\sim 135$ at $Y_p=0.03$.
The distribution for $Y_p=10^{-2}$ exhibits a double-peak structure corresponding to neutron magic numbers $N=50$ and 82.
The spiky features of the QSE abundance distribution at $Y_p=10^{-4}$ become smoother at higher $Y_p$, as repeated cycling of reaction flows suppresses the odd-even effect arising from the nuclear mass differences.

\begin{figure}[htb]
    \centering
    \includegraphics[width=\columnwidth]{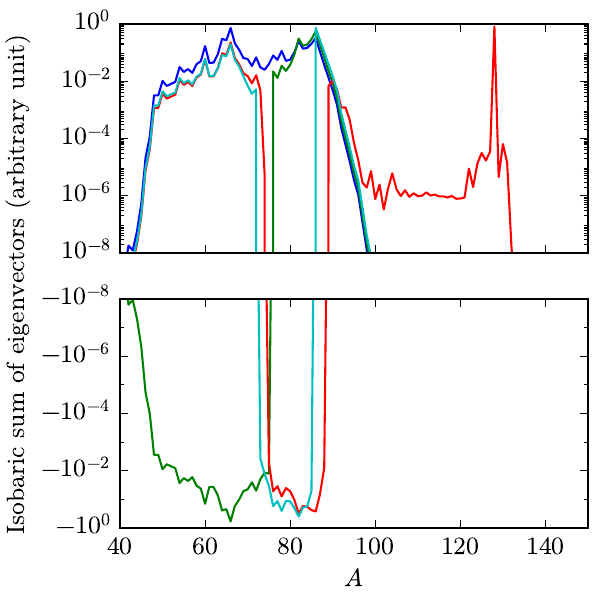}
    \caption{The isobaric sum of four solved eigenvectors as a function of mass number $A$ for the case of $Y_p=10^{-4}$ in Fig.~\ref{fig:sec1_YA}. The blue curve is the NESS solution, while the green, red, and cyan curves correspond to the eigenmodes associated with the next three largest real parts of eigenvalues, respectively.}
    \label{fig:sec1_vec_sum_A}
\end{figure}

For the case of $Y_p=10^{-4}$, we also show the isobaric sum of eigenvectors associated with the next three largest eigenvalues in Fig.~\ref{fig:sec1_vec_sum_A} and compare them to the NESS solution in blue.
Those three eigenmodes clearly exhibit the pattern with a combination of positive and negative regimes because they are transient modes and the total sum is zero as required by Eq.~\eqref{eq:sum_v}.
For example, the eigenmode in red has some positive regimes with mass number $A\sim 120$ and $A\sim 60$, while the sign for $A\sum 80$ is negative.
This suggests that, in a dynamic system where there are finite amount of those nuclei at the initial condition, those nuclei of $A\sim 120$ will be disintegrated toward the nuclei with the mass number $A\sim 80$.

\begin{figure*}[htb]
    \centering
\includegraphics[width=.99\columnwidth]{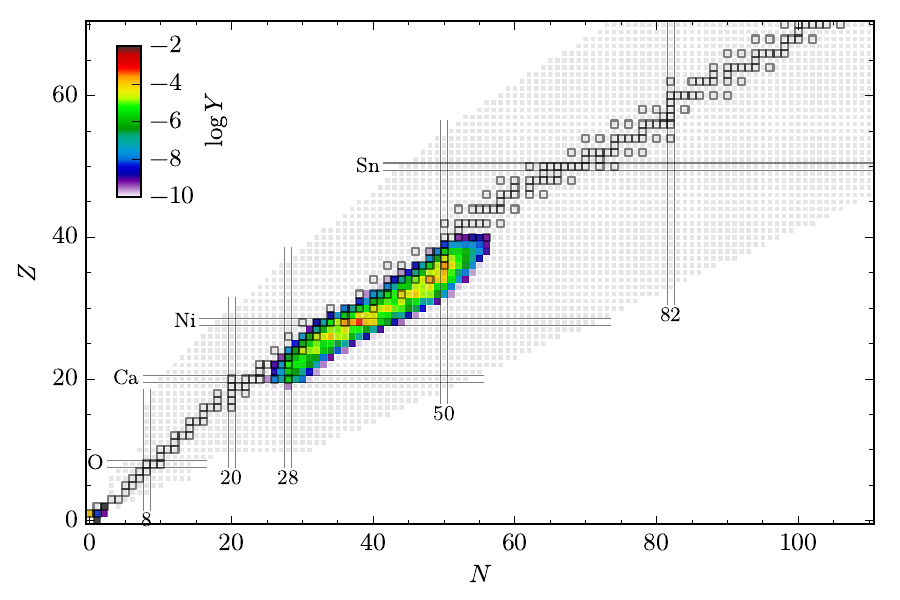}
    \includegraphics[width=.99\columnwidth]{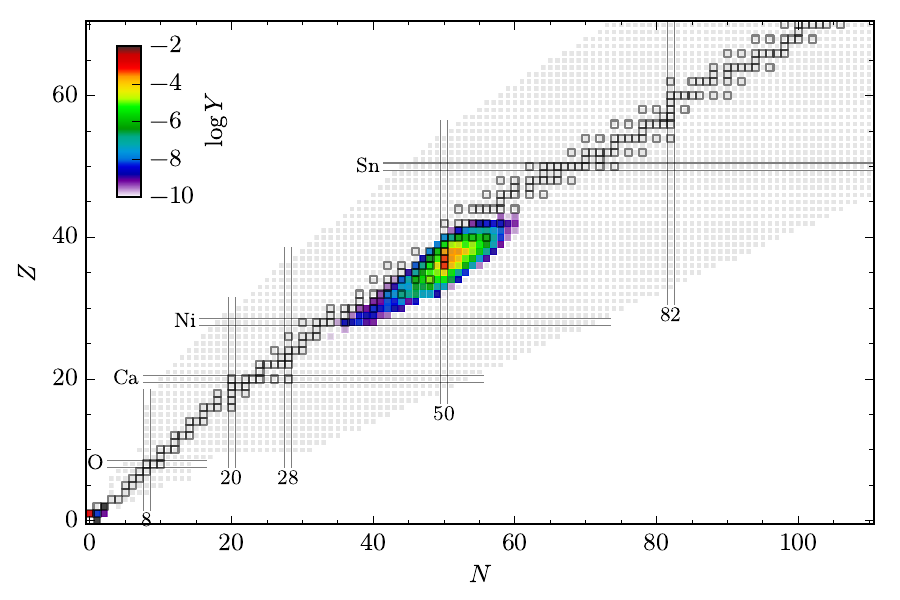}
    \llap{\parbox[b]{11in}{\small $Y_p=10^{-4}$\\\rule{0ex}{1.9in}}}
    \llap{\parbox[b]{4.25in}{\small $Y_p=10^{-3}$\\\rule{0ex}{1.9in}}}
    \includegraphics[width=.99\columnwidth]{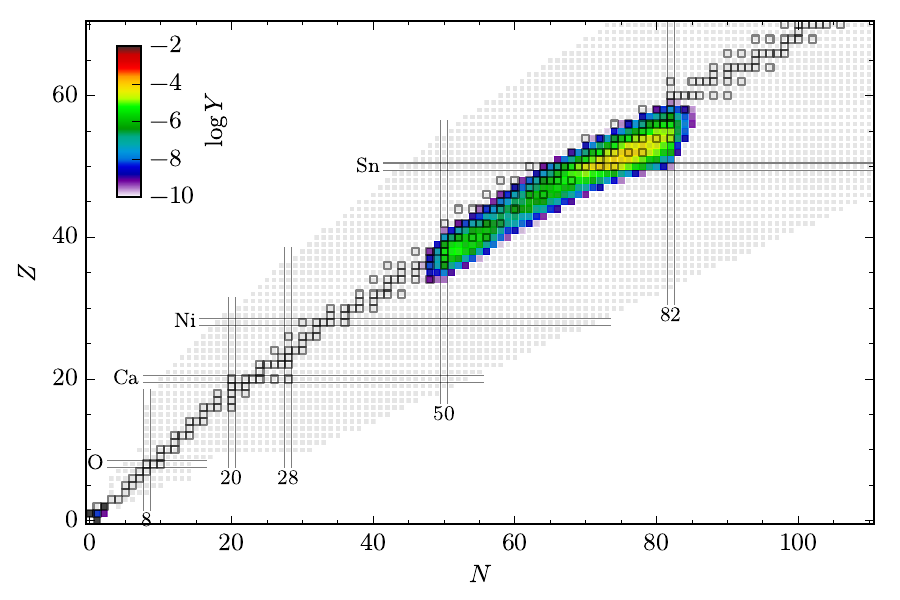}
    \includegraphics[width=.99\columnwidth]{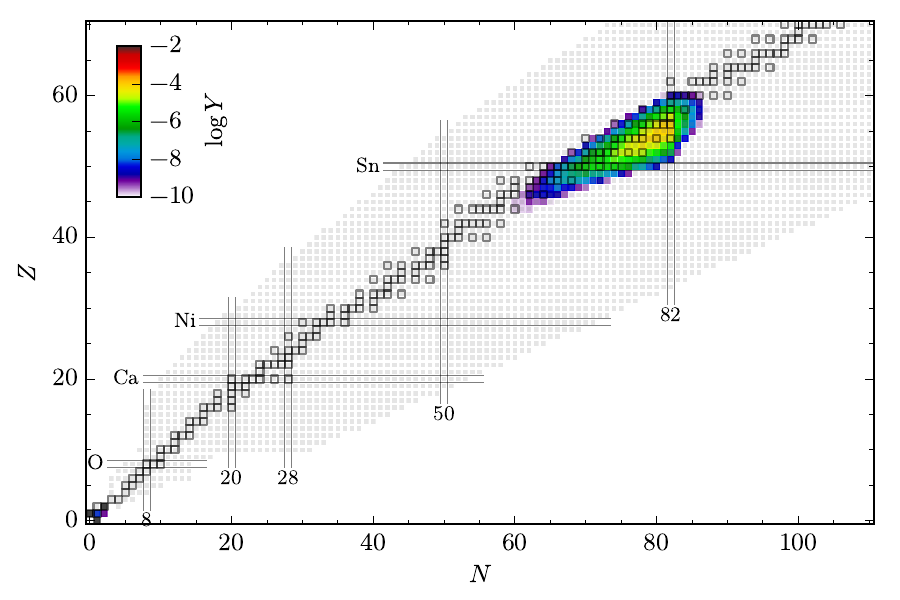}
    \llap{\parbox[b]{11in}{\small $Y_p=10^{-2}$\\\rule{0ex}{1.9in}}}
    \llap{\parbox[b]{4.25in}{\small $Y_p=0.03$\\\rule{0ex}{1.9in}}}
    \caption{Abundance distributions under the nucleon-alpha disequilibrium for four values of $Y_p$. The remaining parameters are same as Fig.~\ref{fig:sec1_YA}.}
    \label{fig:sec1_nzplot}
\end{figure*}

The timescale of the disintegration is related to the eigenvalues of the non-zero real-part eigenmodes.
For this case, the next three largest eigenvalues are $-3.6\times 10^6~\mathrm{s}^{-1}$, $-8.3\times 10^6~\mathrm{s}^{-1}$, and $-2.3\times 10^7~\mathrm{s}^{-1}$, respectively.
This indicates that deviations from the NESS distribution decay rapidly on microsecond timescales.
Among all eigenvalues, most eigenvalues with nonzero real parts are real, while complex eigenvalues appear only beyond approximately the first thousand eigenvalues.
This suggests that the wave-like oscillatory modes are strongly suppressed in the current network, where each nuclear isotope is connected to tens of others.

The differences among the four values of $Y_p$ considered in Fig.~\ref{fig:sec1_YA} are more clearly illustrated in Fig.~\ref{fig:sec1_nzplot}, where the abundance distributions are plotted on the nuclear chart.
For the case with lowest proton abundance, the reaction flow remains concentrated near the $N=50$ isotones, producing the abundance peak around $Z\sim 35$.
As the proton abundance increases, the distribution is shifted progressively toward the $N=82$ region.
The double-peaked abundance distribution in the intermediate case is consistent with that in Fig.~\ref{fig:sec1_YA}.
One peak with relatively smaller abundance is centered near $N=50$, while the other one has a broader expansion over both $Z=50$ and $N=82$ magic numbers.

\begin{figure*}[htb]
    \centering
    \includegraphics[width=.99\columnwidth]{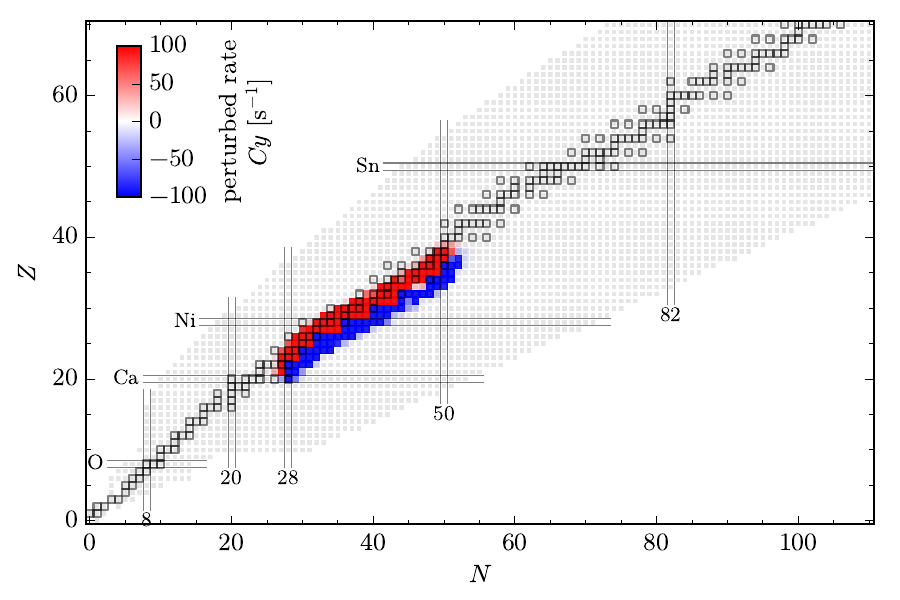}
    \includegraphics[width=.99\columnwidth]{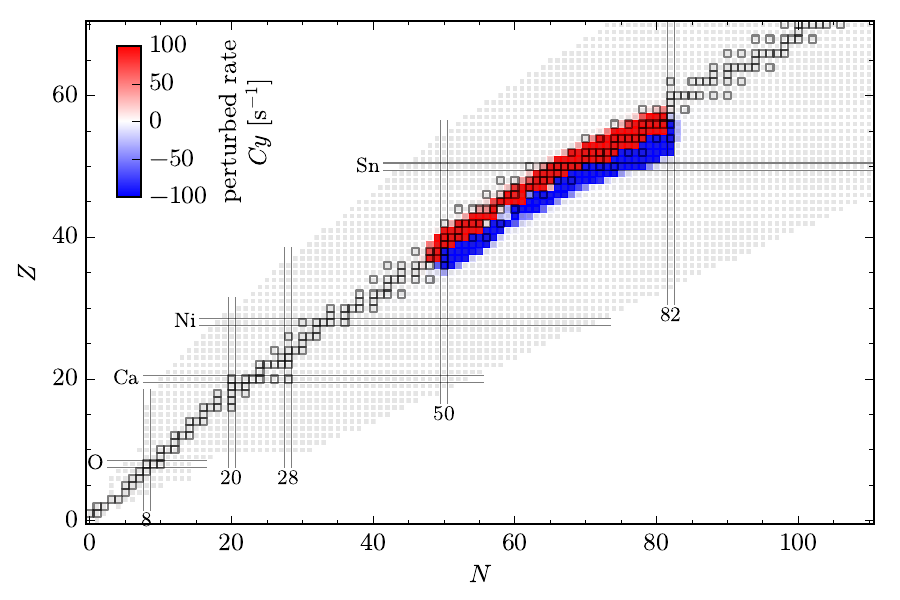}
    \llap{\parbox[b]{10.75in}{\small $Y_p=10^{-4}$\\\rule{0ex}{1.9in}}}
    \llap{\parbox[b]{4in}{\small $Y_p=10^{-2}$\\\rule{0ex}{1.9in}}}
    \includegraphics[width=.99\columnwidth]{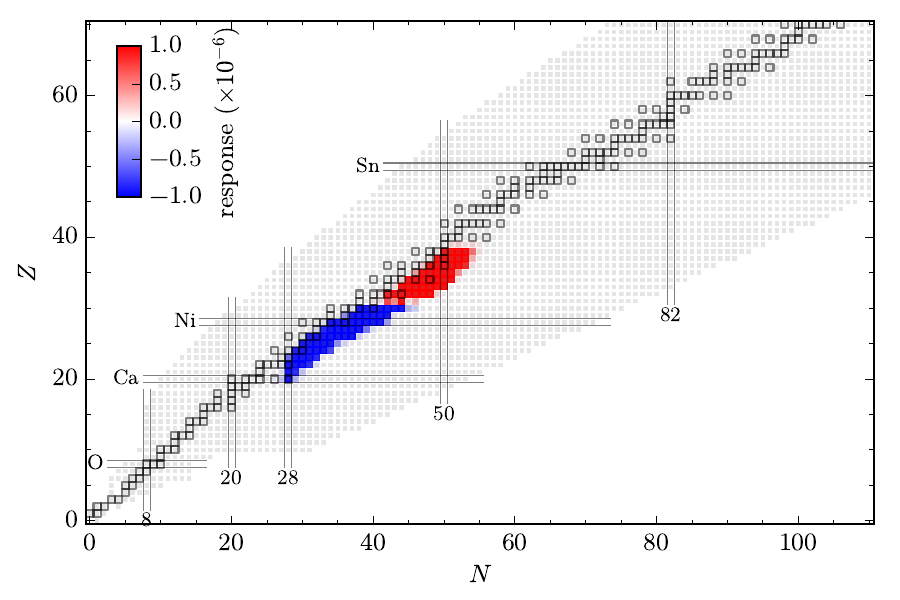}
    \includegraphics[width=.99\columnwidth]{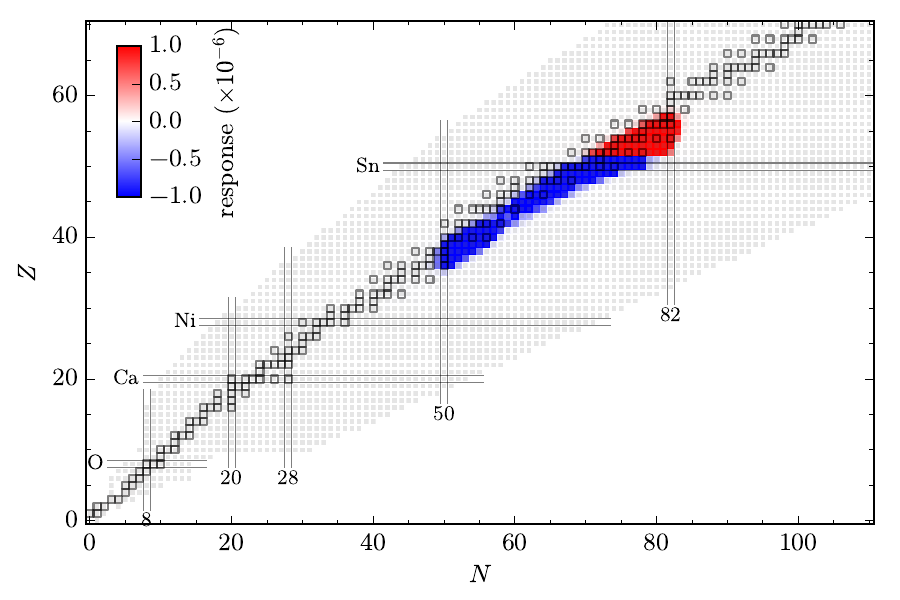}
    \caption{Distributions of perturbed reaction rates (upper panels) and their corresponding responses (lower panels) under the nucleon-alpha disequilibrium for $Y_p=10^{-4}$ (left panels) and $Y_p=10^{-2}$ (right panels). The perturbed reaction channel is $(p,n)$ for all panels. The remaining parameters are same as Fig.~\ref{fig:sec1_YA}.}
    \label{fig:sec1_Cb}
\end{figure*}

To identify the physical origin of this behavior, we apply the perturbation analysis developed in Section~\ref{sec:NESS}.
Since the proton abundance primarily affects the $(p,n)$ reaction channel, we construct the perturbation matrix $\mathbf{C}$ by including only the contributions from these reactions.
By solving Eq.~\ref{eq:Xdyda=-Cy}, the abundance responses $\mathbf y'$ can be obtained.

Figure~\ref{fig:sec1_Cb} shows both the imposed perturbations $\mathbf C \mathbf y$ to the reaction rates and the corresponding abundance responses $\mathbf y'$.
For both $Y_p=10^{-4}$ and $10^{-2}$, an enhancement of the $(p,n)$ reaction rates initially drives the reaction flow toward the $\beta$-stability line.
As this perturbation propagates through the reaction network, however, the overall response is to redistribute the abundances along the stability line toward larger mass numbers.

\begin{figure*}
    \centering
    \includegraphics[width=.99\columnwidth]{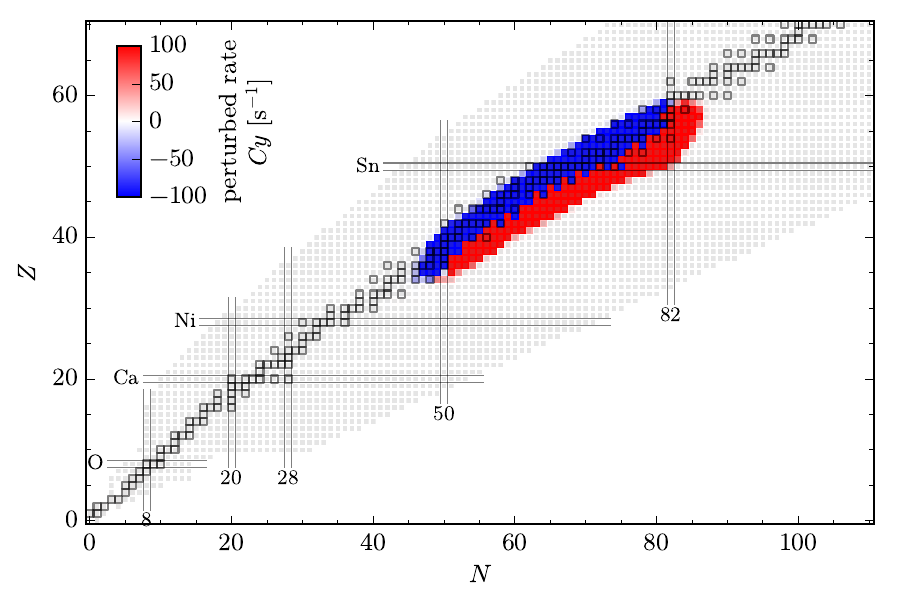}
    \includegraphics[width=.99\columnwidth]{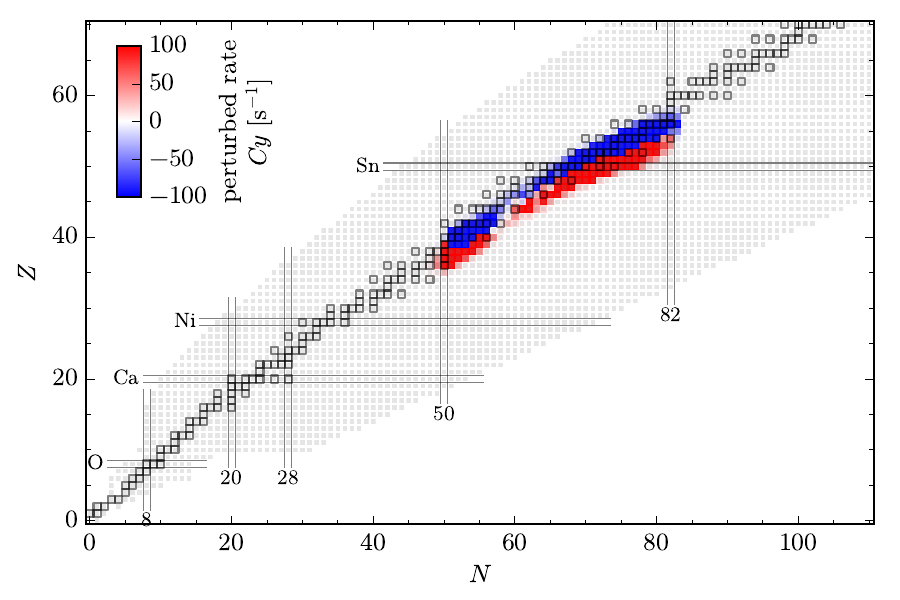}
    \llap{\parbox[b]{10.75in}{\small $(n,\gamma)$\\\rule{0ex}{1.9in}}}
    \llap{\parbox[b]{4in}{\small $(n,\alpha)$\\\rule{0ex}{1.9in}}}
    \includegraphics[width=.99\columnwidth]{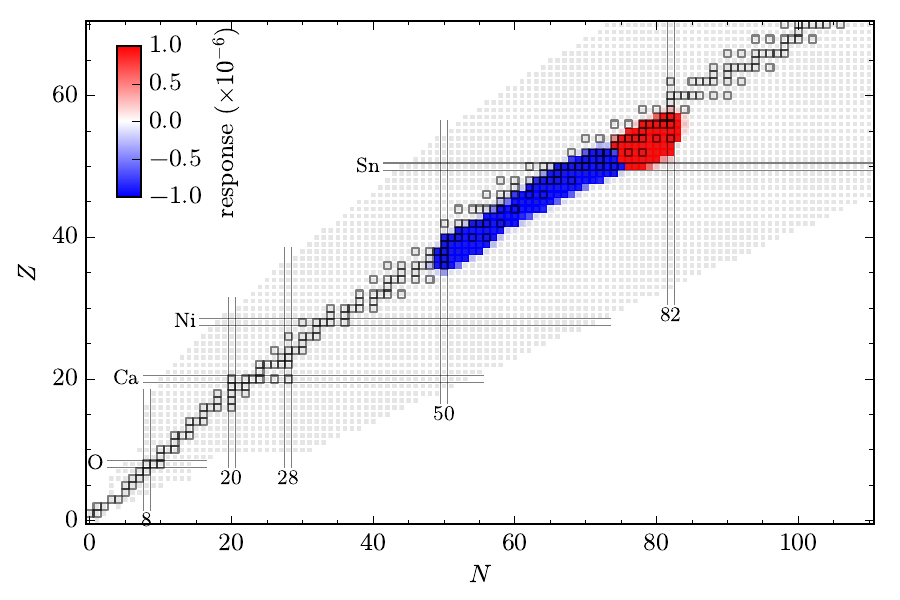}
    \includegraphics[width=.99\columnwidth]{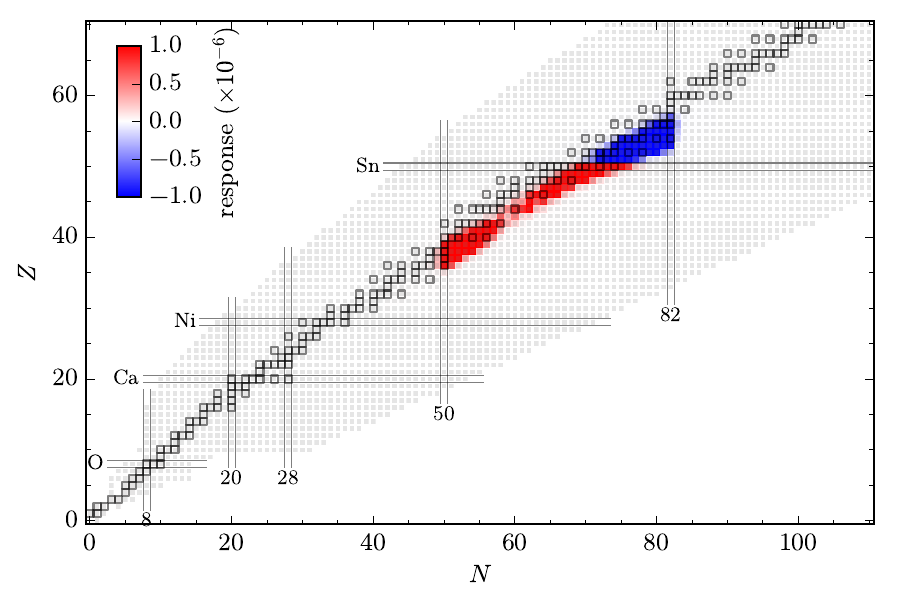}
    \caption{Distributions of perturbed reaction rates (upper panels) and their corresponding responses (lower panels) under the nucleon-alpha disequilibrium with the reaction channels $(n,\gamma)$ (left panels) and $(n,\alpha)$ (right panels) being perturbed. The remaining parameters are given by $\rho=10^6\,\mathrm{g/cm}^3$, $T=5.72$~GK, $Y_n=0.1$, $Y_p=10^{-2}$, and $Y_\alpha\approx 0.173$.}
    \label{fig:sec1_Cb2}
\end{figure*}

A similar analysis can be performed for the neutron-induced reactions by replacing $\mathbf C$ with the perturbation matrix associated with the $(n,\gamma)$ or $(n,\alpha)$ reaction channel for the case with $Y_p=10^{-2}$.
As shown in Fig.~\ref{fig:sec1_Cb2}, the source terms associated with both perturbations drive the abundances toward nuclei with larger neutron numbers.
The propagated responses, however, are qualitatively different.
An enhancement of the $(n,\gamma)$ reaction rates shifts the abundance distribution toward larger mass numbers, whereas an enhancement of the $(n,\alpha)$ reaction rates produces the opposite trend, favoring smaller mass numbers.
Although the two perturbations generate similar local responses, their effects after propagation through the reaction network are fundamentally different.
The responses to perturbations in all other reaction channels, including $(\alpha,\gamma)$, $(\alpha,n)$, $(\alpha,p)$, $(p,\gamma)$, and $(p,\alpha)$, are substantially smaller in magnitude and are therefore not shown.

\begin{figure}
    \centering
    \includegraphics[width=.99\columnwidth]{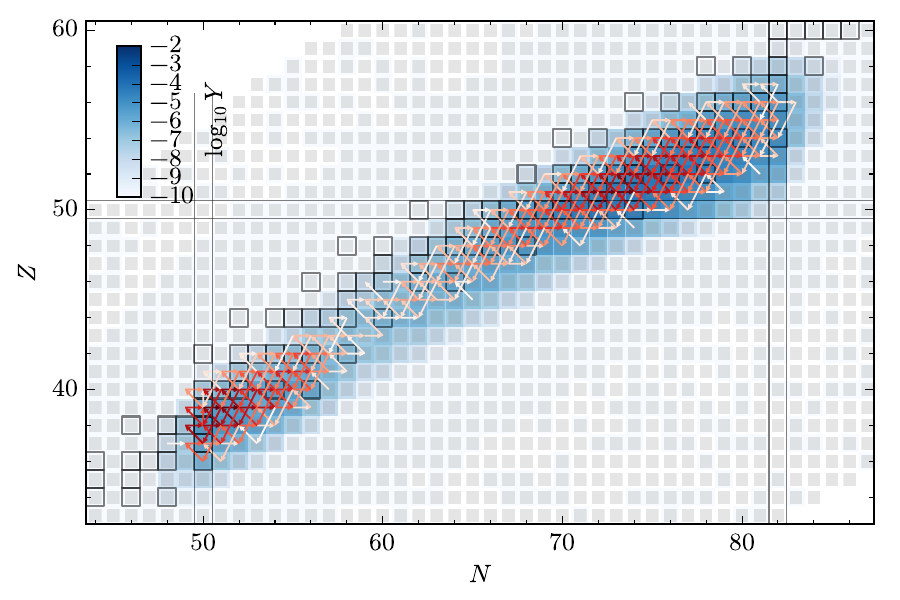}
    \caption{Flux diagram for heavy nuclei under the nucleon-alpha disequilibrium. The color darkness for the arrows indicates the flux influence from white ($64\,\mathrm{s}^{-1}$) to red ($6.4\times 10^3\,\mathrm{s}^{-1}$). The remaining parameters are same as Fig.~\ref{fig:sec1_Cb2}.}
    \label{fig:sec1_flux_diagram}
\end{figure}

We further examine the reaction fluxes in Fig.~\ref{fig:sec1_flux_diagram}.
The dominant reaction flow forms a highly overlapping cyclic current consisting of the sequence $(p,n)$--$(n,\gamma)$--$(n,\gamma)$--$(p,n)$--$(n,\gamma)$--$(n,\alpha)$.
Within this cycle, the $(p,n)$ and $(n,\gamma)$ reactions cooperatively drive the reaction flow toward heavier nuclei, whereas the $(n,\alpha)$ reaction acts in the opposite direction by returning material to lower mass numbers.
Consequently, although perturbations to the $(n,\gamma)$ and $(n,\alpha)$ channels both initially favor neutron-rich nuclei locally, they modify opposite branches of the dominant reaction cycle.
Therefore, the resulting abundance responses differ qualitatively after propagating through the network, illustrating that the global NESS structure, rather than the local reaction direction alone, determines the final outcome.

\begin{figure*}
    \centering
    \includegraphics[width=.99\columnwidth]{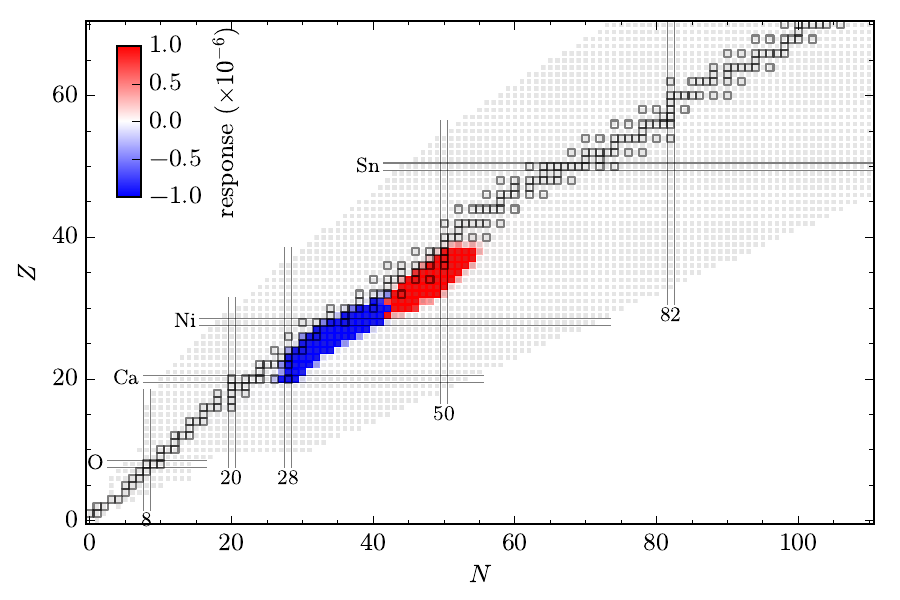}
    \includegraphics[width=.99\columnwidth]{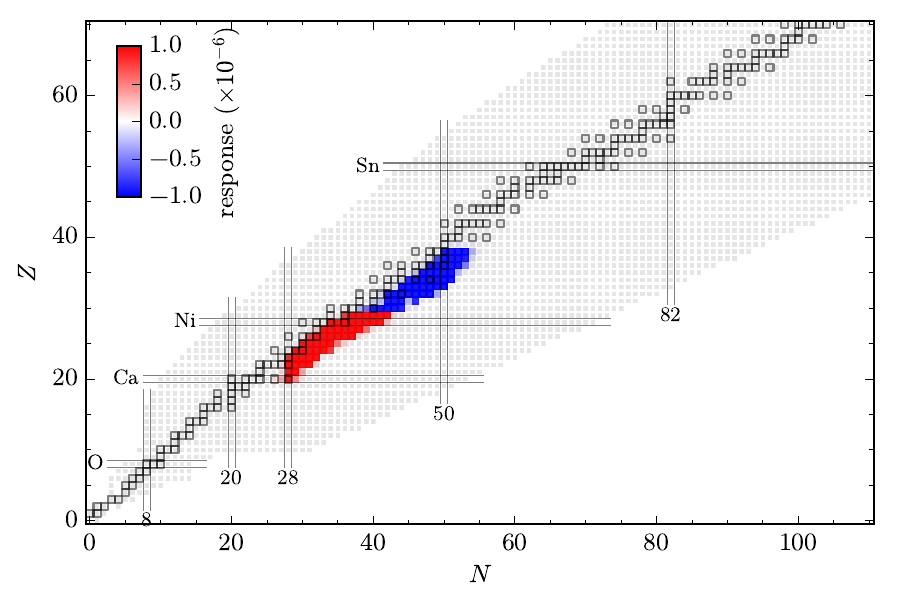}
    \llap{\parbox[b]{10.75in}{\small $(n,\gamma)$\\\rule{0ex}{1.9in}}}
    \llap{\parbox[b]{4in}{\small $(n,\alpha)$\\\rule{0ex}{1.9in}}}
    \includegraphics[width=.99\columnwidth]{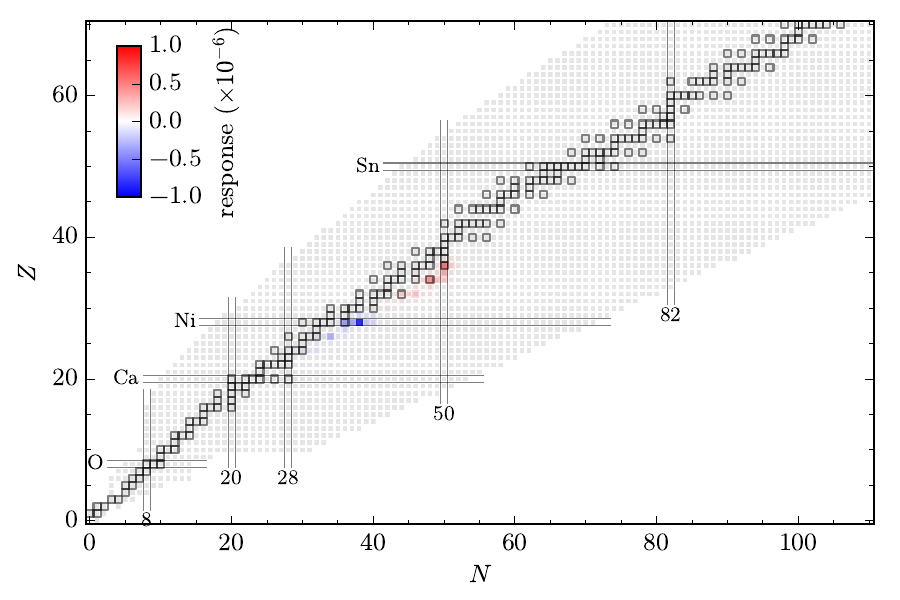}
    \includegraphics[width=.99\columnwidth]{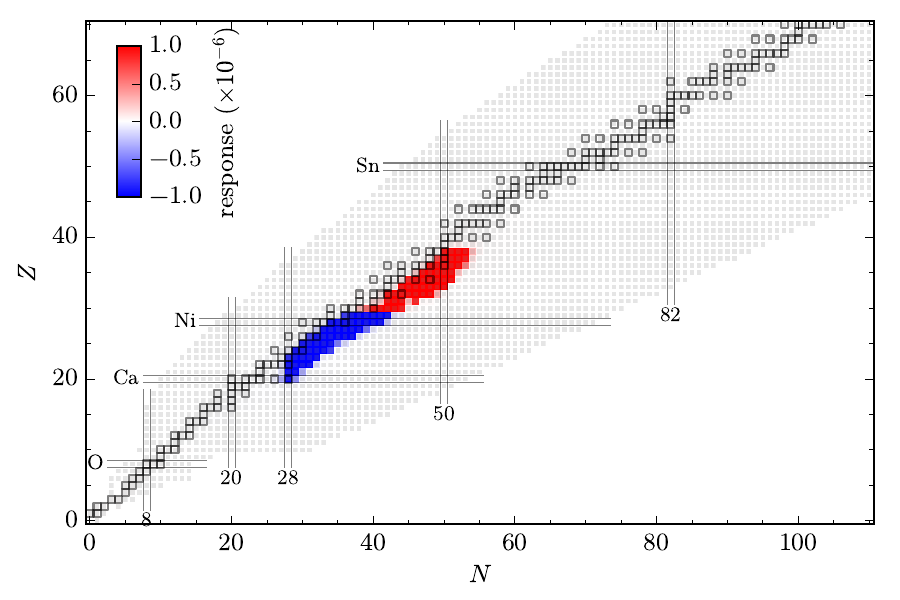}
    \llap{\parbox[b]{10.75in}{\small $(\alpha,\gamma)$\\\rule{0ex}{1.9in}}}
    \llap{\parbox[b]{4in}{\small $(\alpha,n)$\\\rule{0ex}{1.9in}}}
    \caption{Abundance responses for the QSE distribution with four reaction channels being perturbed. The remaining parameters are same as Fig.~\ref{fig:sec1_Cb2} except $Y_p=10^{-4}$.}
    \label{fig:sec1_Cb3}
\end{figure*}

The disequilibrium case with $Y_p=10^{-2}$ does not show strong sensitivity to $\alpha$-particle capture, but it may depend on the NESS distribution itself.
For example, Fig.~\ref{fig:sec1_Cb3} shows the abundance responses $\mathbf y'$ for the QSE distribution obtained with $Y_p=10^{-4}$.
In this case, the NESS distribution lies predominantly to the left of $N=50$ neutron magic-number line, and the $(\alpha,n)$ reaction rates are enhanced relative to those for $Y_p=10^{-2}$.
Consequently, the responses to $(\alpha,n)$ reactions become comparable in magnitude to those associated with $(n,\gamma)$ and $(n,\alpha)$ reactions.
Other channels, such as $(\alpha,\gamma)$, make negligible contribution.

\subsection{Steady States of $\beta$-Flow}
During the main phase of the rapid neutron-capture process, neutron captures and photo-disintegration are typically much faster than $\beta$ decay, establishing local equilibrium within each isotopic chain.
The nucleosynthesis flow toward heavier nuclei is then regulated primarily by $\beta$ decay, leading to the classical $\beta$-flow steady state provided that the $r$-process duration is longer than the individual $\beta$-decay rates.
This regime provides one of the earliest and most important examples of a steady state in nuclear astrophysics \cite{cameron1983steady,cameron1983waiting}.

We consider both hot and cold $r$-process conditions, with the heavy-nuclei cluster comprising all isotopes that can undergo $\beta^-$ decay and satisfy $10<N<160$ and $10<Z<80$.
We include the following reactions $(n,~\gamma)$, $(\gamma,~n)$, and $\beta^-$ decay.
No neutrino interactions are included.
The parameters for hot $r$-processes are $\rho=256\,\mathrm{g/cm}^3$, $T=0.47$~GK, $Y_n=1.2\times 10^{-2}$, and $n_n=1.9\times 10^{24}~\mathrm{cm}^{-3}$.
The parameters for cold $r$-processes are $\rho=4.33\,\mathrm{g/cm}^3$, $T=0.15$~GK, $Y_n=1.2\times 10^{-2}$, and $n_n=3.2\times 10^{22}~\mathrm{cm}^{-3}$.
The above conditions are actually taken from two parametric $r$-process nuclear network calculations with the setup present in Appendix~\ref{sec:parametric}.
The hot case has the characteristic values of $Y_e=0.2$, $S\approx 24~k_B$/nucleon, $\tau_{\rm exp}=10$~ms at 5~GK.
The cold case has the characteristic values of $Y_e=0.2$, $S\approx 10~k_B$/nucleon, $\tau_{\rm exp}=1$~ms at 5~GK.
The steady state solution for both hot and cold cases is evaluated when the neutron-to-seed ratio $n/s$ is close to 1 in the nuclear network calculation where abundances are dynamically evolved.

To construct a NESS for the reaction network, the network must form a closed system with no terminal nuclei.
In the present calculation, the physical reaction network terminates at the heaviest nucleus included in the model and therefore does not admit a stationary solution.
To remove this singularity, we introduce an artificial reaction that returns the terminal nucleus with the largest mass number to the nucleus with smallest mass number, thereby closing the reaction network and ensuring the existence of a NESS.
The rate of this artificial reaction is chosen to be sufficiently large as $10^{10}\,\mathrm{s}^{-1}$ that it does not influence the reaction flow within the physical region of interest, serving only to provide the boundary condition required for the steady-state calculation.

\begin{figure*}
    \centering
    \includegraphics[width=.99\columnwidth]{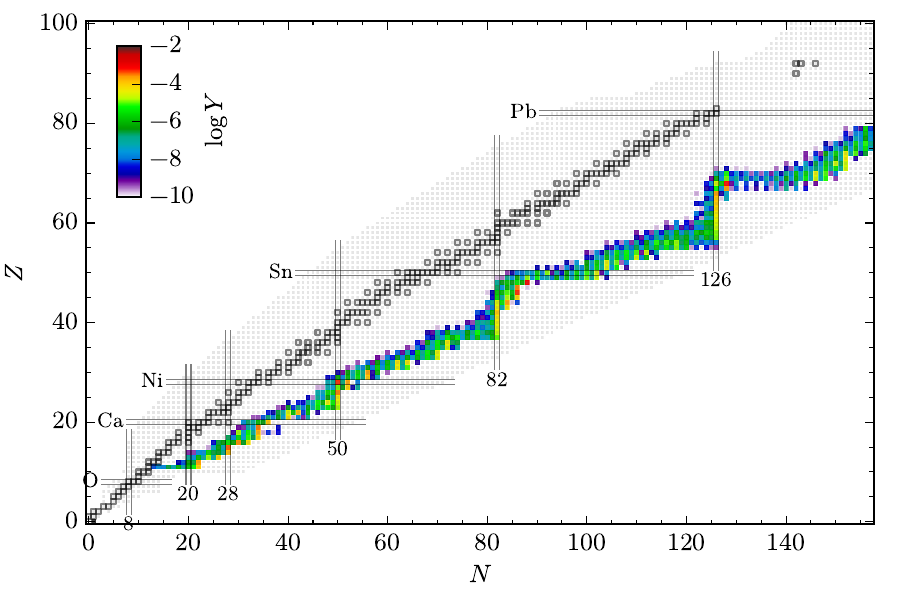}
    \includegraphics[width=.99\columnwidth]{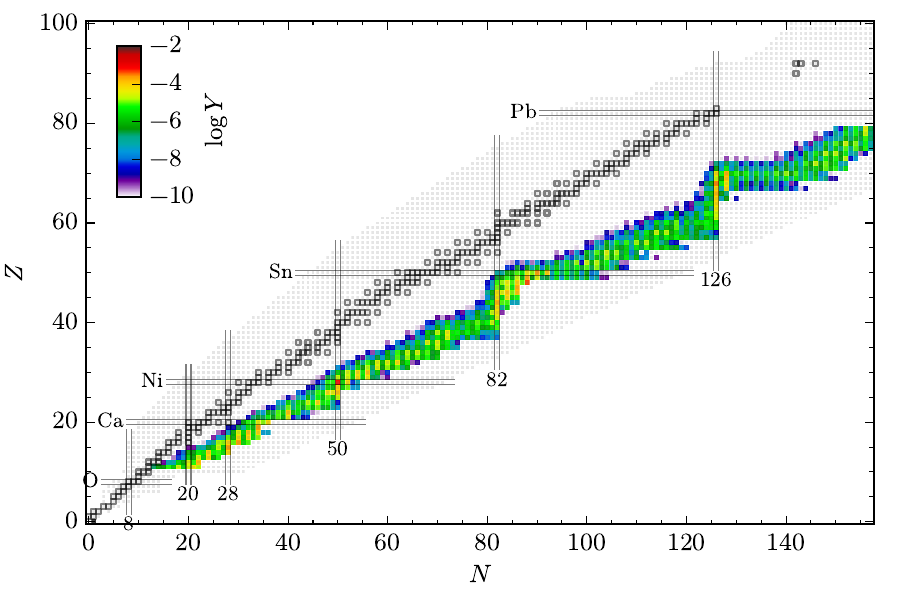}
    \llap{\parbox[b]{10.75in}{\small hot\\\rule{0ex}{1.9in}}}
    \llap{\parbox[b]{4in}{\small cold\\\rule{0ex}{1.9in}}}
    \includegraphics[width=.99\columnwidth]{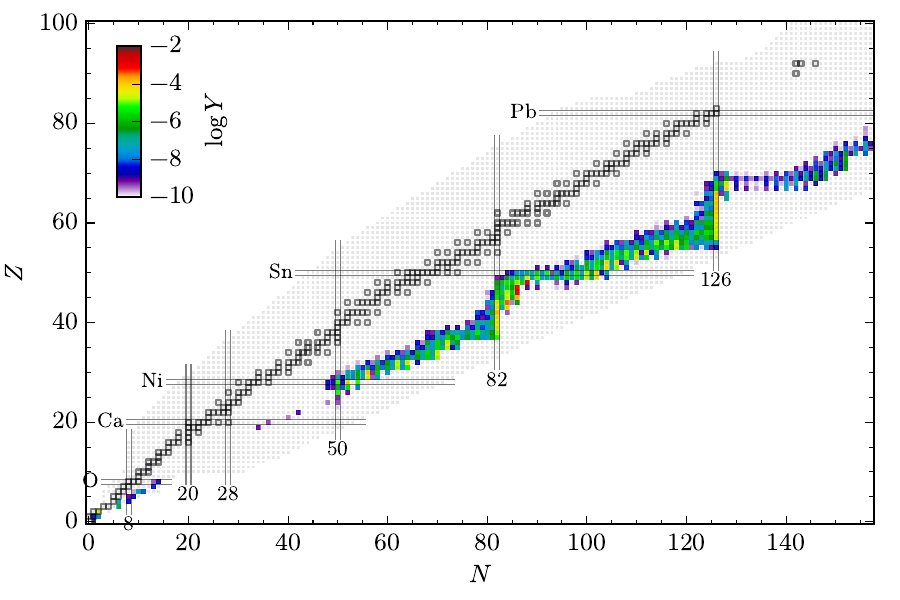}
    \includegraphics[width=.99\columnwidth]{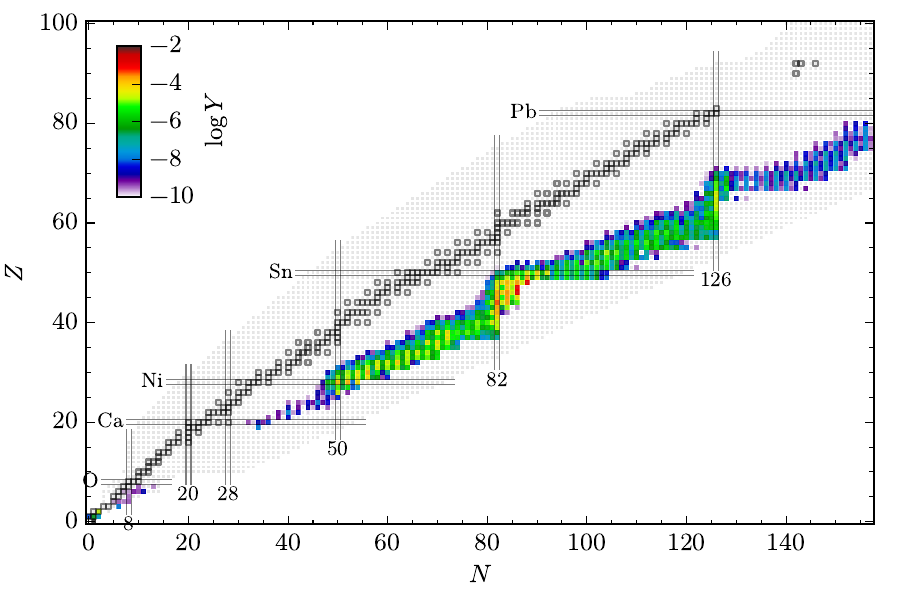}
    \caption{Abundance distributions in $N$--$Z$ plane of the $\beta$-flow steady-state solution (upper panels) and from the nuclear network calculations evaluated when the neutron-to-seed ratio $n/s$ is close to 1 (lower panels) for hot (left panels) and cold (right panels) $r$-processes. The conditions for hot $r$-processes are $T=0.47$~GK and $n_n=1.9\times 10^{24}~\mathrm{cm}^{-3}$. The conditions for cold $r$-processes are $T=0.15$~GK and $n_n=3.2\times 10^{22}~\mathrm{cm}^{-3}$. Protons and $\alpha$ particles are negligible.}
    \label{fig:sec2_nzplot}
\end{figure*}

\begin{figure*}
    \centering
    \includegraphics[width=.97\columnwidth]{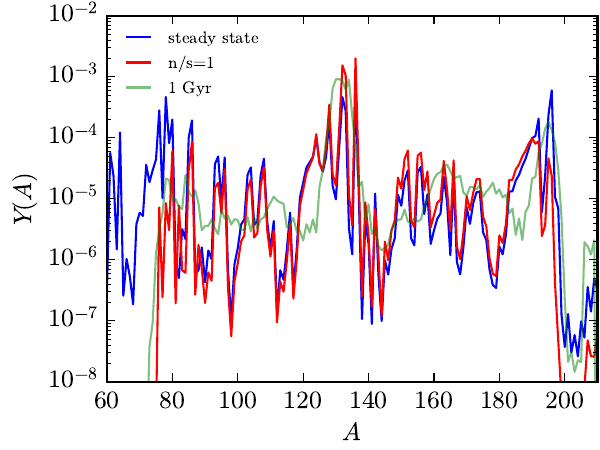}
    \includegraphics[width=.97\columnwidth]{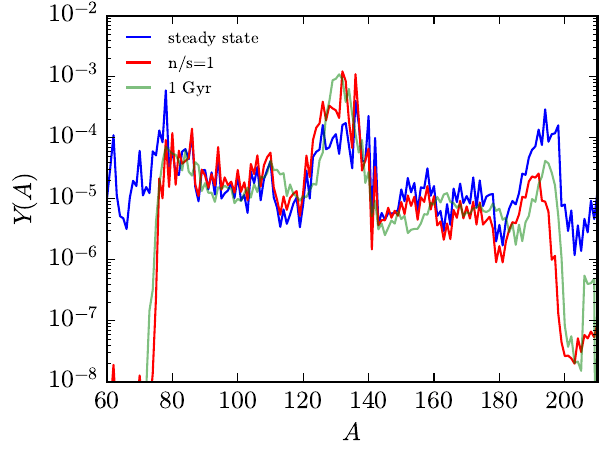}
    \llap{\parbox[b]{9in}{\small hot\\\rule{0ex}{2.1in}}}
    \llap{\parbox[b]{2in}{\small cold\\\rule{0ex}{2.1in}}}
    \includegraphics[width=.97\columnwidth]{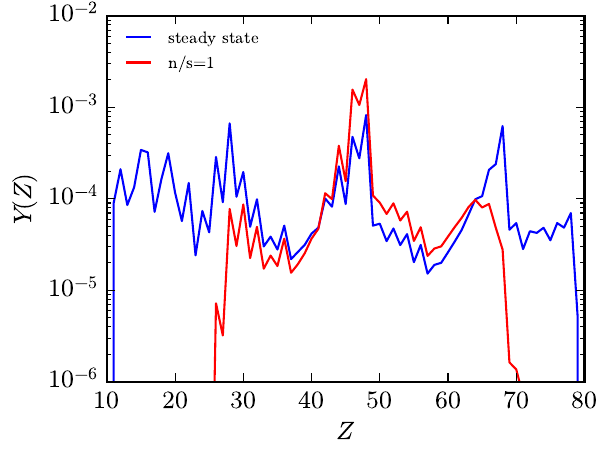}
    \includegraphics[width=.97\columnwidth]{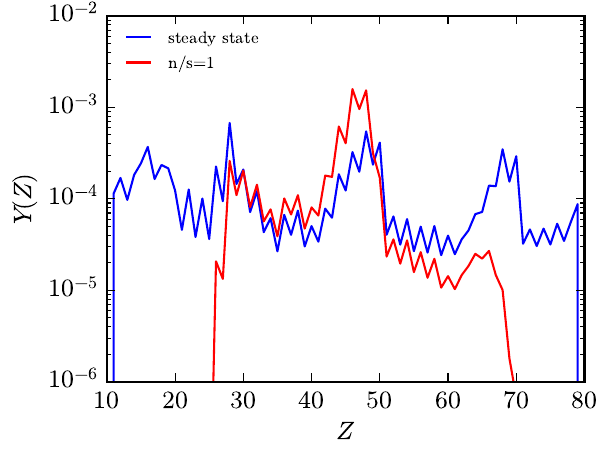}
    \caption{Abundance distributions of the $\beta$-flow steady-state as a function of mass number $A$ (upper panels) and atomic number $Z$ (lower panels) for hot (left panels) and cold (right panels) $r$-processes. The NESS solutions (blue) are compared to the nuclear network results evaluated at $n/s=1$ (red) and at 1~Gyr (green). The parameters are the same as in Fig.~\ref{fig:sec2_nzplot}.}
    \label{fig:sec2_YA}
\end{figure*}

Although both hot and cold $r$-processes exhibit the same $\beta$-flow NESS, the underlying equilibrium are different as shown in Fig.~\ref{fig:sec2_nzplot}.
In the hot $r$-process, neutron captures are counteracted predominantly by photo-disintegration within each isotopic chain before the slower $\beta$ decays transport material to the next element.
In contrast, photo-disintegration becomes weaker in the cold $r$-process, and the steady state is established directly between neutron captures and $\beta$ decays.
In both cases, the distributions in the $N$--$Z$ plane predicted by the NESS generally agree with that obtained from the full reaction-network calculation.
Both the overall reaction path and the abundance distribution are reproduced, indicating that the network has evolved sufficiently close to the $\beta$-flow NESS under these conditions.

Figure~\ref{fig:sec2_YA} shows both isobaric and elemental abundance distributions.
The results from network calculations show wave-like propagation, with the abundance pattern spanning over a narrower range of $Z$ than the NESS solutions.
Nuclei with $Z\lesssim 25$ have already been depleted, while those with $Z\gtrsim 65$ have not yet been produced.
Within the intermediate range, the NESS solutions agree well with dynamical network calculations in both cases.
In particular, the agreement is strongest for the isobaric abundance distributions in hot $r$-process among all other panels, owing to the imposed chemical equilibrium along each isotopic chain.
The isobaric abundance distribution exhibits much larger oscillations in the hot $r$-process than in the cold $r$-process because the neutron separation energies shape $(n,\gamma)$--$(\gamma,n)$ equilibrium.
In the latter process, photodisintegration becomes weaker, and the competition is primarily between neutron captures and $\beta$ decays.
Correspondingly, the abundance distribution at $n/s=1$ is more similar to the final distribution at 1~Gyr in the cold $r$-process than in the hot $r$-process.

The elemental abundance panels also show the agreement between the NESS solutions and the dynamical calculations.
The dynamical calculations show slightly higher abundances in the central region than in the wings of the distributions, owing to the relatively low neutron-capture rates near the $N=82$ neutron shell closure.

\begin{figure*}
    \centering
    \includegraphics[width=.32\linewidth]{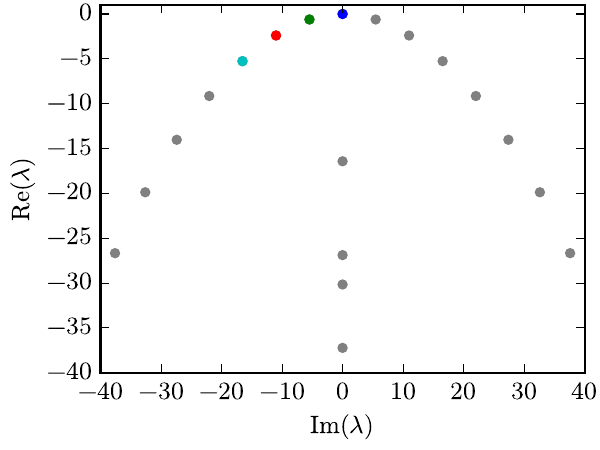}
    \includegraphics[width=.32\linewidth]{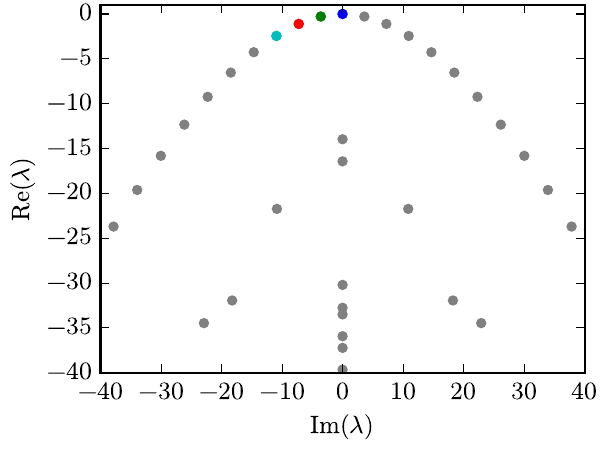}
    \llap{\parbox[b]{8in}{\small hot\\\rule{0ex}{1.5in}}}
    \llap{\parbox[b]{3.4in}{\small cold\\\rule{0ex}{1.5in}}}
    \includegraphics[width=.32\linewidth]{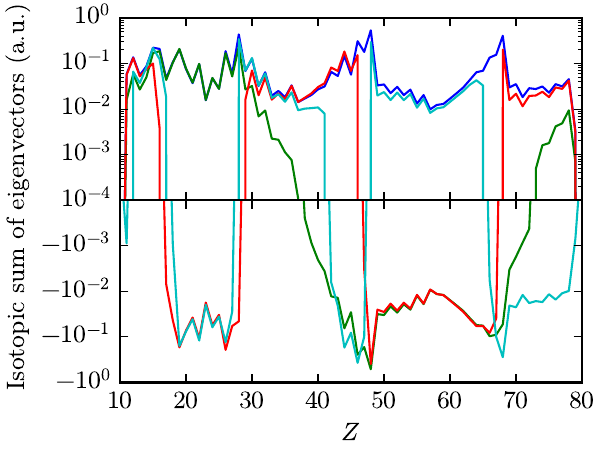}
    \caption{Distributions of eigenvalues in complex plane for hot (left panel) and cold (middle panel) $r$-processes.
    The unit of both axes is $\mathrm{s}^{-1}$.
    The right panel shows the isotopic sum of real parts of four eigenvectors that have the largest real parts of eigenvalues. The colors of the curves correspond to those labeled in the left panel. The parameters for hot and cold $r$-processes are the same as in Fig.~\ref{fig:sec2_nzplot}.}
    \label{fig:sec2_eigenvalues_complex_plane}
\end{figure*}

These features of the dynamical results can be understood from a different perspective, if we take the distribution as the combination of all possible eigenvectors instead of merely the steady-state mode.
Figure~\ref{fig:sec2_eigenvalues_complex_plane} compares the eigenvalues distributions for both cases in complex plane.
Both hot and cold $r$-process spectra contain a series of complex-conjugate eigenvalue pairs extending from the steady-state eigenvalue at the origin.
The eigenvalue spectrum differs in the scenario of nucleon-alpha disequilibrium, where most eigenvalues are real and negative.
Both cases also show purely decaying modes with real eigenvalues starting at $\mathrm{Re}(\lambda)\sim -15~{\rm s}^{-1}$.

In cold $r$-process, an additional component of complex eigenvalues lies between the predominant complex series and the purely decaying modes, indicating that the absence of $(n,\gamma)$--$(\gamma,n)$ equilibrium allows more degrees of freedom.
The real parts of the corresponding eigenvalues are approximately $-20~{\rm s}^{-1}$, implying relaxation timescales of order $0.1~$s and requiring a longer time for these modes to decay substantially, provided the environment conditions remain approximately unchanged over this interval.

The predominance of the complex series in hot $r$-process reflects the presence of flowing structure of the network.
The corresponding eigenmodes exhibit multipole-like patterns associated with the loop configuration aided by the artificial closure of the $\beta$-flow.
The leading complex modes, shown in green in the right panel of Fig.~\ref{fig:sec2_eigenvalues_complex_plane}, exhibits one region with a negative real part in the isotopically summed eigenvector.
The next two modes, shown in red and cyan, exhibit two and three such regions, respectively.
For longer expansion timescales, higher-order oscillatory modes have more time to decay, leaving only the lower-order modes with appreciable amplitudes and allowing the abundance distribution to spread over a wider range of atomic numbers.
As given from the estimation in Appendix~\ref{sec:parametric}, the expansion timescales in both cases are around 0.1~s. A a result, the leading complex modes can survive while higher-multiple modes decay rapidly, accounting for the good agreement between NESS solutions and network results within $25<Z<65$.

Our results demonstrate that the NESS solution can provide an accurate description of the abundance distribution without requiring a full dynamical network calculation.
Furthermore, the eigenvalue spectrum offers insight into the underlying structure of the reaction network, particularly its cyclic features and associated wave-like eigenmodes.
The decay timescales of the leading non-steady-state eigenmodes govern the persistence of these modes and provide insight into how the width of the abundance distribution depends on the expansion timescale.

\subsection{Neutrino-Assisted Steady State in $\nu r$-Process}
The $\nu r$-process may occur in neutron-rich environments with intense neutrino irradiation, where neutrino absorption plays a central role in the nuclear reaction network.
Unlike the classical $r$-process, the influx brought by neutron captures are counteracted primarily by neutrino absorption rather than by photodisintegration or $\beta$ decay after the neutron freeze-out.
The reaction flow is therefore governed by a distinct combination of reactions while maintaining a quasi-stationary state among the dominant processes.
This unique state modifies both the reaction pathways and the resulting abundance evolution, making the $\nu r$-process an ideal example for examining the NESS framework.

\begin{figure*}
    \centering
    \includegraphics[width=.99\columnwidth]{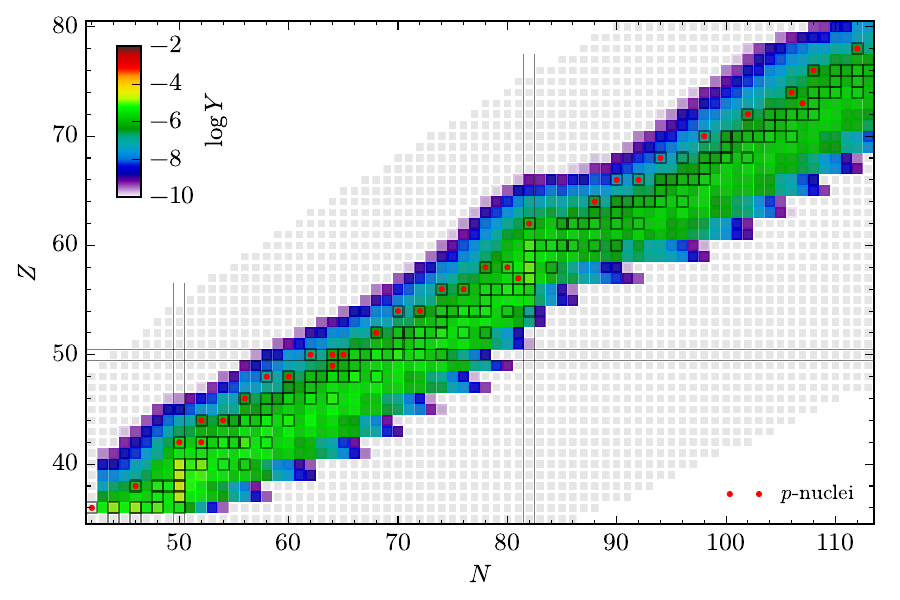}
    \includegraphics[width=.99\columnwidth]{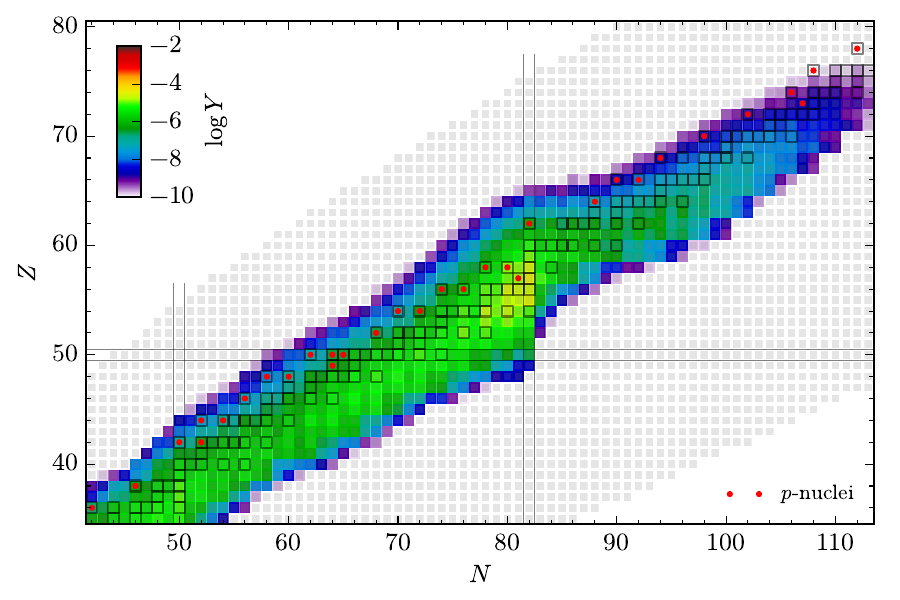}
    \caption{Abundance distributions in $N$--$Z$ plane for the steady-state solution (left panel) and from the nuclear network calculations of the case III in Ref.~\cite{xiong2024production} (right panels) for the steady-state solution in $\nu r$-process. The conditions are evaluated at $T=1.7$~GK. Heavy-lepton-flavor neutrinos are not included. The $p$-nuclei are labeled by the red dots.}
    \label{fig:sec3_nzplot}
\end{figure*}

\begin{figure*}
    \centering
    \includegraphics[width=.99\columnwidth]{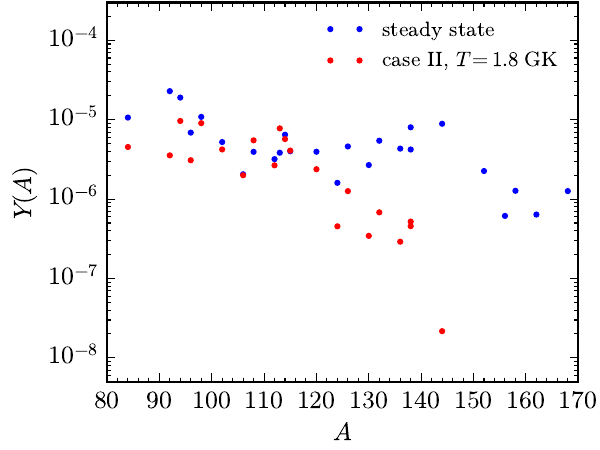}
    \includegraphics[width=.99\columnwidth]{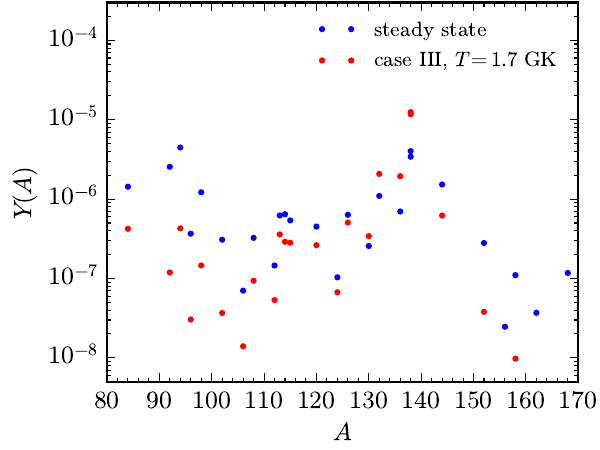}
    \caption{Abundance distributions of the $p$-nuclei as a function of mass number $A$ for the conditions evaluated in case II (left panels) and case III (right panels) of the nuclear network calculations in Ref.~\cite{xiong2024production}. Heavy-lepton neutrinos are not included.
    }
    \label{fig:sec3_YA}
\end{figure*}

\begin{figure}
    \centering
    \includegraphics[width=.99\columnwidth]{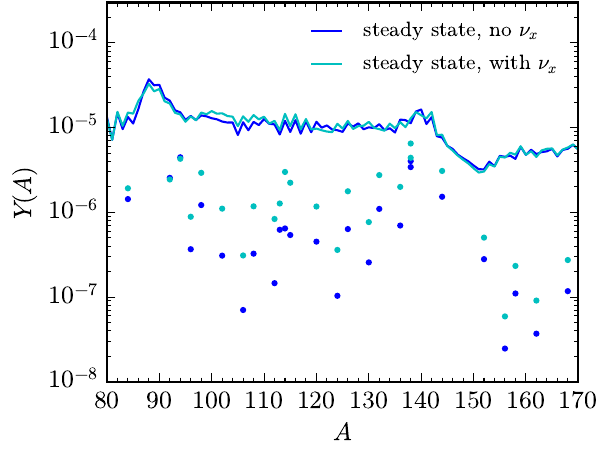}
    \caption{NESS abundance distributions a function of mass number $A$ for the conditions evaluated in case III of the nuclear network calculations in Ref.~\cite{xiong2024production}. Solid curves are for the isobaric sum, and dots are for $p$-nuclei. Heavy-lepton neutrinos $\nu_x$ are included in the cases of cyan color, as compared to the case without $\nu_x$ in blue. The conditions are given in the main text.}
    \label{fig:sec3_YA_nux}
\end{figure}

In this model the heavy-nuclei cluster include all isotopes which can undergo $\beta^-$ decay and within the condition of $42<N<122$ and $35<Z<85$ .
We include $(n,~\gamma)$ reactions, $\beta^+$ decay, and both charged-current and neutral-current neutrino absorption.
As in the $\beta$-flow steady-state scenario, we add an additional reaction with the rate of $10^{10}~\mathrm{s}^{-1}$ from the terminal nucleus to the starting nucleus to remove this singularity of the network.
The conditions are taken at $T\sim 2~$GK from two parametric trajectories corresponding to cases II and III in Ref.~\cite{xiong2024production}.
For case II, we use $\rho=1.5\times 10^4\,\mathrm{g/cm}^3$, $T=1.8$~GK, $Y_n=6.6\times 10^{-10}$, and set $\nu_e$ flux as $f_{\nu_e} = 2\times 10^{42}\, \mathrm{cm}^{-2}\,\mathrm{s}^{-1}$.
For case III, the parameters are $\rho=1.1\times 10^4\,\mathrm{g/cm}^3$, $T=1.7$~GK, $Y_n=4.5\times 10^{-9}$, $f_{\nu_e} = 3.4\times 10^{42}\, \mathrm{cm}^{-2}\,\mathrm{s}^{-1}$.
Protons and $\alpha$ particles are negligible because their impact is relatively smaller than $(n,\gamma)$ reactions on the domain of neutron number $N>50$.
The mean energy for $\nu_e$'s is $\sim 15.75$~MeV.
We do not consider antineutrinos.
For heavy-lepton-flavor neutrinos, we first consider the cases without them, and then an example on the effects of heavy-lepton neutrinos will be shown with the flux of $3.4\times 10^{42}\, \mathrm{cm}^{-2}\,\mathrm{s}^{-1}$ and the mean energy of $31.5$~MeV.

Figure~\ref{fig:sec3_nzplot} presents the abundance distributions in $N$--$Z$ plane for the $\nu r$-process in case III.
The NESS solution in the left panel is compared to the network result in the right panel.
Unlike the classical $r$-process, the reaction flow follows a path much closer to the valley of $\beta$ stability and, in some regions, even extends beyond it as a result of the competition between neutrino absorption and neutron captures.
The resulting abundance distribution occupies an essentially broader region of the nuclear chart.
The broader abundance distribution arises because neutrino absorption exhibits a much weaker dependence on neutron number than photodisintegration.
Consequently, the steady state between neutron captures and neutrino absorption is maintained over a wider range of isotopes, allowing the NESS to extend across a larger portion of each isotopic chain.
This behavior contrasts with the $r$-process, where the strong neutron-number dependence of the photodisintegration rates confines the abundance distribution to a much narrower region.

A similarity to the $r$-process is that, nuclei are produced from the seed at low mass number, and the network is also in a flowing configuration.
The network result shows more concentrated abundance at around $N=82$ magic number, while the nuclei heavier than this line have not been produced in essential amount.

The abundances of the synthesized $p$-nuclei are compared in Fig.~\ref{fig:sec3_YA} for the two cases.
In general, the patterns match each other well for the nuclei with $A<120$ in case II and $A<140$ in case III.
In particular, the steady-state calculation reproduces both the relative abundances and the characteristic variation among the individual $p$-nuclei in case III.

The results presented above are calculated without including the absorption of heavy-lepton neutrinos.
We now examine the sensitivity of the abundance distributions to the heavy-lepton neutrinos with the flux and mean energy given at the beginning of this section.
Figure~\ref{fig:sec3_YA_nux} compares the isobaric abundance distributions and the $p$-nuclei abundances calculated with and without $\nu_x$ for case III.
Including heavy-lepton neutrino absorption does not significantly change the isobaric distribution, but it substantially enhances the abundance of $p$-nuclei, as neutrino-induced spallation shifts the nuclear abundance distributions toward the proton-rich side.

Once again, we find that the abundance patterns produced under the influence of neutron capture and neutrino absorption in $\nu r$ process can be well approximated by the NESS solutions.
The same NESS analysis can also be use to investigate the sensitivity of the abundance distributions to the neutrino flux and individual neutrino absorption channels.

\subsection{Neutron-Rich $r$-Process with Sufficient Fission Cycles}
In extremely neutron-rich environments, the reaction flow can proceed to nuclei sufficiently heavy to undergo fission.
The resulting fission fragments are recycled through subsequent neutron captures, producing repeated fission cycles that strongly influence the final abundance distribution.
The continuous recycling of material through fission provides another important example in which the reaction network may be understood from the perspective of NESSs.

In this model the heavy-nuclei cluster include all isotopes which can undergo $\beta^-$ decay and within the condition of $75<N<189$, $36<Z<100$, and $5N>9(Z-10)$.
The set of reactions consists of $(n,~\gamma)$, $(\gamma,~n)$, $\beta^-$ decay, spontaneous, $\beta$-delayed, and neutron-induced fissions.
No neutrino interactions are included.
In this case, the addition artificial reaction is not need because the fission automatically close the network reaction flow.
We take one parametric trajectory with $Y_e=0.05$, $S\approx 24~k_B$/nucleon, $\tau_{\rm exp}=10$~ms at 5~GK.
Two time snapshots are used, one when the neutron-to-seed ratio $n/s$ is 10, and the other at $n/s=1$.
The conditions at $n/s=10$ are $\rho=4.8\,\mathrm{g/cm}^3$, $T=0.2$~GK, $Y_n=6.66\times 10^{-2}$, and the neutron density $n_n=1.9\times 10^{23}~\mathrm{cm}^{-3}$.
The conditions at $n/s=1$ are $\rho=3.87\,\mathrm{g/cm}^3$, $T=0.197$~GK, $Y_n=6.3\times 10^{-3}$, and $n_n=1.5\times 10^{22}~\mathrm{cm}^{-3}$.
Protons and $\alpha$ particles are negligible.

\begin{figure*}
    \centering
    \includegraphics[width=.99\columnwidth]{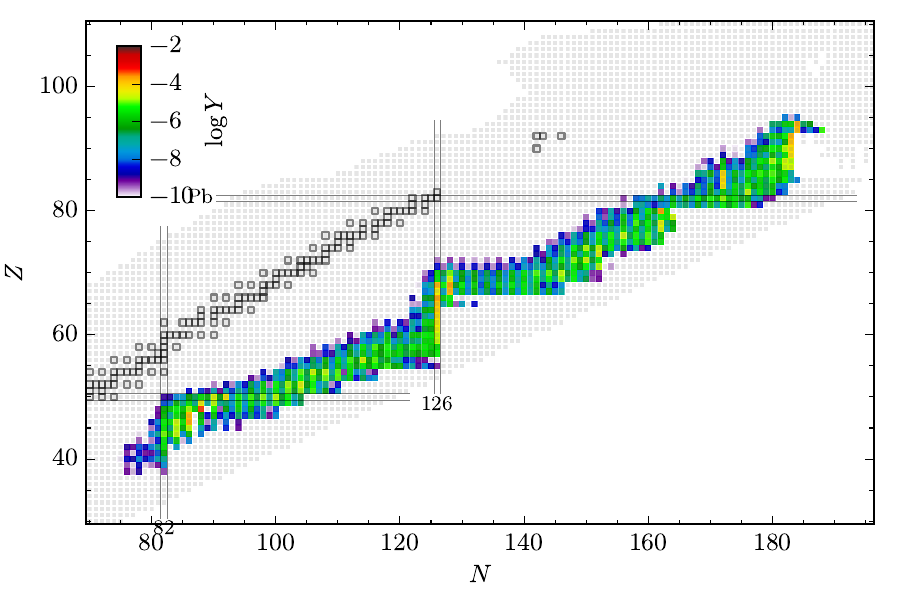}
    \includegraphics[width=.99\columnwidth]{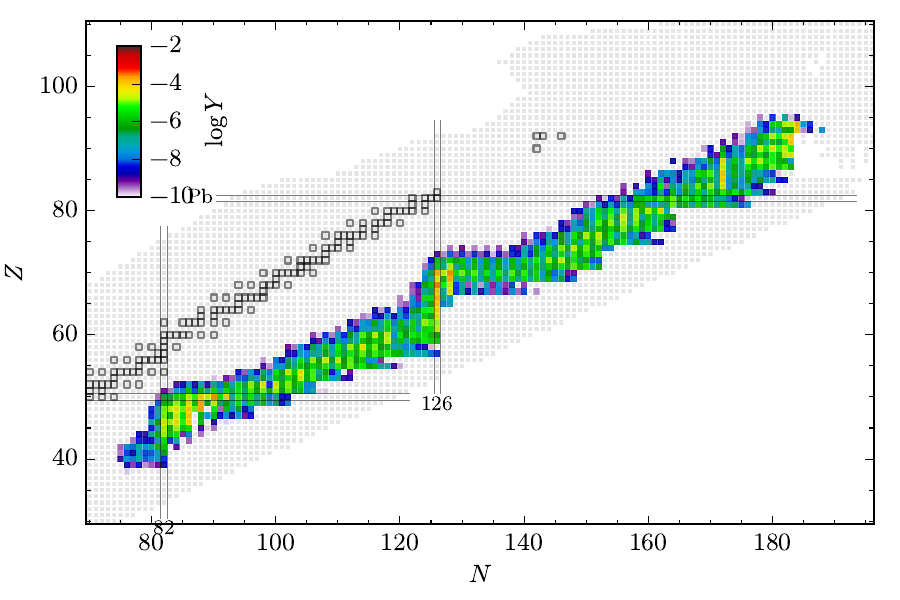}
    \includegraphics[width=.99\columnwidth]{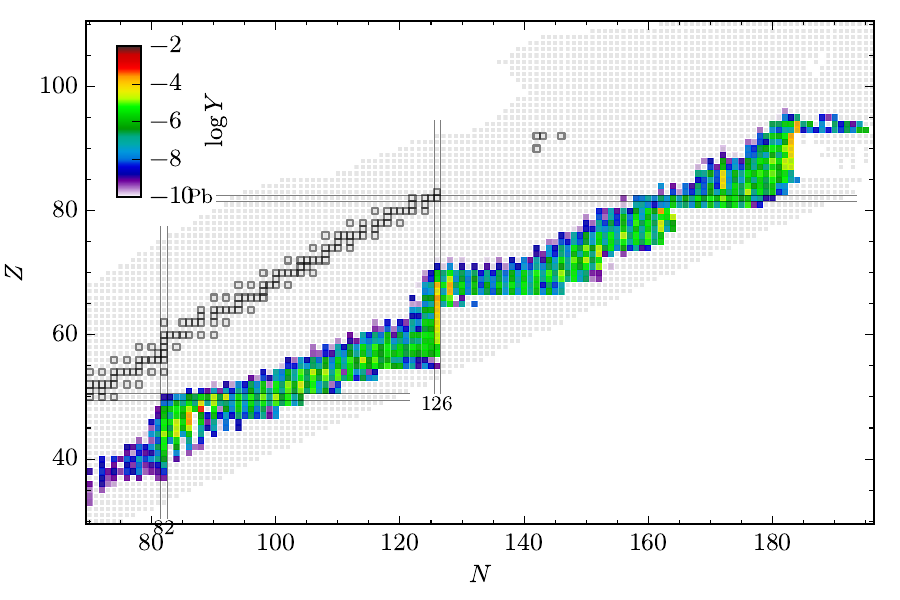}
    \includegraphics[width=.99\columnwidth]{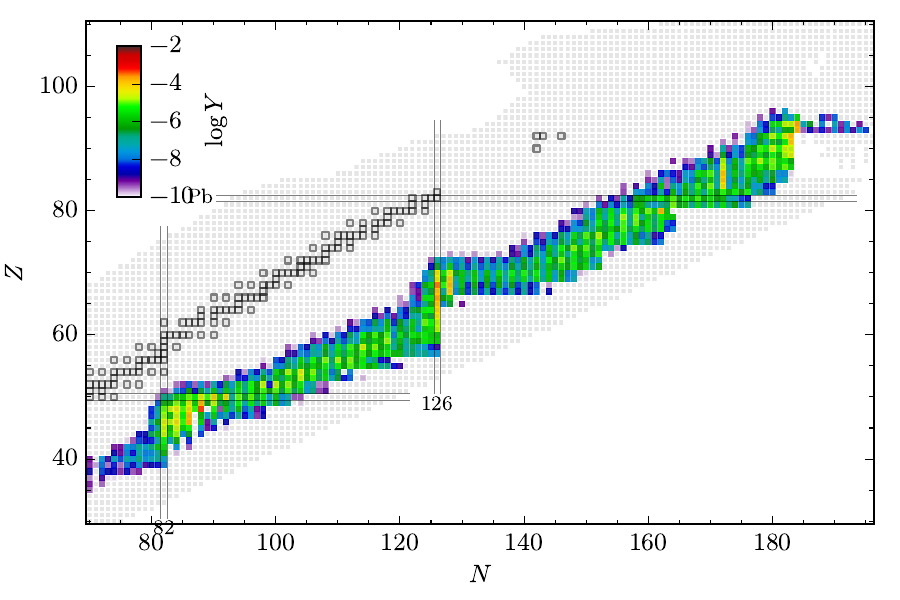}
    \includegraphics[width=.99\columnwidth]{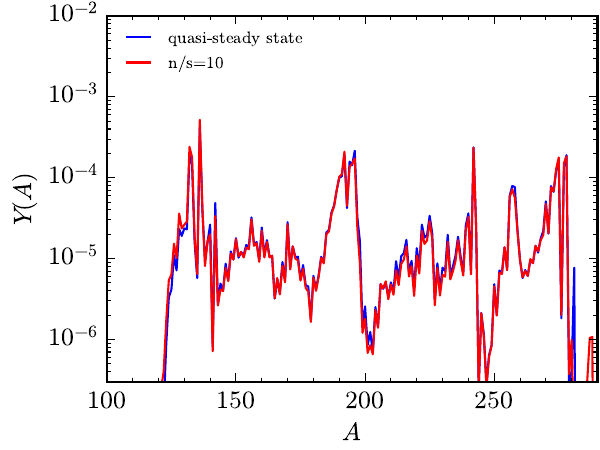}
    \includegraphics[width=.99\columnwidth]{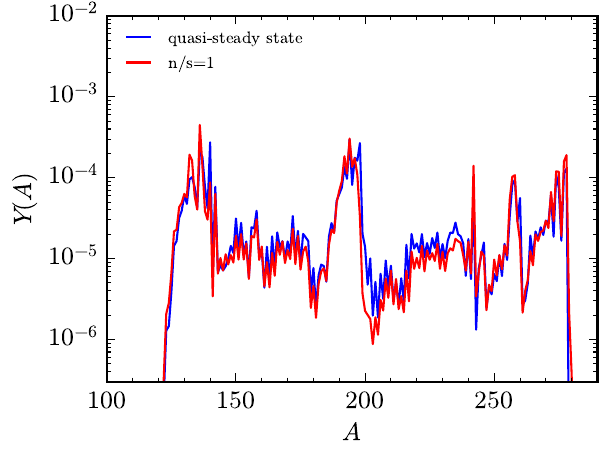}
    \caption{Abundance distributions in $N$--$Z$ plane of the QSS solution (upper panels), from the nuclear network calculations (middle panels), and isobaric abundance distributions of the QSS solution (lower panels) at two time stages with the neutron-to-seed ratio $n/s=10$ (left panels) and $n/s=1$ (right panels) for sufficiently neutron-rich $r$-process. The conditions at $n/s=10$ are $T=0.2$~GK and $n_n=1.9\times 10^{23}~\mathrm{cm}^{-3}$. The conditions at $n/s=1$ are $T=0.197$~GK and $n_n=1.5\times 10^{22}~\mathrm{cm}^{-3}$. Protons and $\alpha$ particles are negligible.}
    \label{fig:sec4_nzplot}
\end{figure*}

Figure~\ref{fig:sec4_nzplot} compares the QSS solutions with the corresponding nuclear-network calculation at two evolutionary stages mentioned above.
For both stages, the steady-state solution reproduces the overall reaction path and abundance distribution, indicating that the heavy nuclei remain close to a QSS throughout the fission-cycling phase.

As the neutron-to-seed ratio decreases from 10 to 1, the neutron-capture rates become progressively weaker, leading to a larger deviation between the QSS solution and the network calculation.
Nevertheless, the overall agreement remains good even at $n/s=1$, demonstrating that the QSS approximation continues to capture the dominant features of the abundance distribution.

The good agreement can be understood from the nature of the fission process itself.
Unlike ordinary reaction channels, a single fission event populates a broad distribution of daughter nuclei rather than a single reaction product.
Consequently, each fission cycle redistributes abundance over many nuclei, increasing the branching of the reaction flow and preventing material from remaining localized along a single reaction path.
This continual redistribution smooths the abundance distribution and makes the QSS approximation less sensitive to local deviations from equilibrium, allowing it to reproduce the results of the full reaction-network calculation with good accuracy.

Because the fission does not conserve the total abundance, the eigenvalue we obtained during the diagonalization for the dominant mode is not zero.
This can give us an interesting indication on how fast the heavy sector consumes neutrons during the fission cycle.
In the above two snapshots, the dominant eigenvalues $\lambda_0$ are 0.99~s$^{-1}$ and 0.62~s$^{-1}$ at $n/s=10$ and 1, respectively.
Taking average, it is approximately 0.8~s$^{-1}$.
From the neutron abundance, it can be calculated that the amplification factor of the heavy abundance is $6.4$\%.
This amplification factor requires a time duration of 0.08~s, which is consistent with the time difference from the nuclear network calculation where the first stage is at $t=1.21$~s, and the second at $t=1.30$~s.

\section{Discussions and Conclusion}
\label{sec:conclusion}
In this work, we developed a general theoretical framework for describing NESSs in nucleosynthesis within a reduced linear description of the relevant subset of the nuclear reaction network.
We formulated the steady-state condition as an equilibrium between reaction flows and external source and sink terms.
This formulation extends the concepts of NSE and QSE to systems in which chemical equilibrium in detailed balance is replaced by a steady circulation of reaction fluxes.
The resulting framework provides a unified description of steady abundance distributions in non-equilibrium reaction networks and establishes a systematic approach for analyzing nucleosynthesis beyond equilibrium conditions.

We applied this framework to four representative nucleosynthesis scenarios spanning different steady-state regimes.
The case involving disequilibrium between nucleons and $\alpha$ particles demonstrated how the NESS formalism describes the abundance evolution after the breakdown of QSE during $\alpha$-rich freeze-out.
The classical $\beta$-flow steady state showed that the traditional steady-state picture is naturally recovered as a special case of the general NESS framework.
The $\nu r$-process illustrated that the same framework reproduces the characteristic abundance pattern of $p$-nuclei when neutron captures match primarily neutrino absorption rather than photodisintegration or $\beta$ decay.
Finally, the fission-cycling phase of the $r$-process demonstrated that the QSS approximation remains accurate when a dominant growing mode fixes the relative abundance distribution while the total abundance changes.
Together, these examples show that the present framework unifies a broad class of nucleosynthesis phenomena that have traditionally been treated separately and provides a systematic basis for understanding steady-state behavior in nuclear reaction networks.

Beyond the applications considered here, this framework provides a new perspective on the production of heavy nuclei under non-equilibrium conditions.
Rather than viewing nucleosynthesis merely as a sequence of individual reactions or equilibrium states, the NESS formalism emphasizes the collective organization of reaction flows over the network.
This perspective is particularly important for understanding nucleosynthesis in astrophysical environments in which heavy nuclei are synthesized through reaction pathways that are not readily described within the conventional equilibrium framework.
Perturbation theory also provides a systematic tool for exploring the sensitivity of nuclear abundance distributions to individual reaction channels.

The formulation developed in this work opens new directions for studying nucleosynthesis through graph-based methods such as spectral clustering, network centrality, sensitivity analysis, and coarse-graining to identify dominant reaction pathways and construct reduced reaction networks \cite{newman2018networks}.
The framework can also be extended to analyze reaction networks that incorporates nuclear isomeric states alongside their corresponding ground states \cite{misch2021astromers}.
More broadly, interpreting nucleosynthesis as transport on a weighted directed graph provides new tools for understanding how heavy nuclei are synthesized through the collective dynamics of complex reaction networks.

\begin{acknowledgments}
We thank Heamin Ko, Yong-Zhong Qian, Friedel Thielemann, and Meng-Ru Wu for helpful discussions.
ZX acknowledges support of the European Research Council (ERC) under the ERC Grant (NeuTrAE, No. 101165138).
GMP acknowledges ERC support under the European Union’s Horizon 2020 research and innovation program (ERC Advanced Grant KILONOVA No. 885281).
Additional support is acknowledged from the the Deutsche Forschungs-gemeinschaft (DFG, German Research Foundation) – Project-ID 279384907 – SFB 1245 (subproject B06) and  Project-ID MA 4248/3-1.
The work is partially funded by the European Union. Views and opinions expressed are however those of the author(s) only and do not necessarily reflect those of the European Union or the European Research Council Executive Agency. Neither the European Union nor the granting authority can be held responsible for them.
We acknowledge the following software: \textsc{matplotlib}~\cite{matplotlib}, \textsc{numpy}~\cite{numpy}, and \textsc{scipy}~\cite{scipy}.
\end{acknowledgments}

\bibliographystyle{apsrev4-1}
\bibliography{references_260309.bib}

\appendix
\section{Parametric $r$-Process Trajectory}
\label{sec:parametric}
To compare the NESS results with full $r$-process network calculations, we carry out a parametric study of nucleosynthesis.
The reaction network has been employed in previous studies, e.g., Refs.~\cite{just2023endtoend,just2025rhine}.
Each calculation starts at time $t=0$ from an initial state with temperature $T_0=10$~GK, density $\rho_0$, and electron fraction $Y_{e,0}$.
The subsequent density evolution is prescribed by
\begin{equation}
    \rho(t) = \begin{cases}
        \rho_0 e^{-t/\tau}, & t\leq t_{\rm tr}, \\
        \rho_0 e^{-t_{\rm tr}/\tau} \left(\frac{t-t_{\rm tr}}{3 \tau}+1 \right)^{-3}, & t>t_{\rm tr},
    \end{cases}
\end{equation}
where $\tau$ is a characteristic expansion timescale, and $t_{\rm tr}=3(1-\ln 0.6)\tau$ is the transition time between the two expansion regimes, inspired by Ref.~\cite{lippuner2015rprocess}.
We assume NSE between $T_0$ and $T=8$~GK, and for $T<8$~GK we use the full reaction network to evolve the nuclear composition.
The corresponding temperature is evolved together with the thermodynamic response to this energy release and to the expansion, following \citet{mendoza2015nuclear}.
Consequently, $T(t)$, $Y_e(t)$, and the entropy $S(t)$ are evolved self-consistently together with $\rho(t)$ at each time step.

Each parametric trajectory is characterized by the expansion timescale 
\begin{equation}
    \tau_{\rm exp}=\left|\frac{d \ln \rho(t)}{dt}\right|^{-1} = \begin{cases}
        \tau, & t\leq t_{\rm tr}, \\
        \frac{t-t_{\rm tr}}{3}+\tau, & t>t_{\rm tr},
    \end{cases}
\end{equation}
together with the values of $Y_e$ and $S$ at $T=5$~GK.

\end{document}